\documentclass[epj]{svjour}
\usepackage{graphics}

\begin{document}
\title{Model-independent data reductions of elastic proton-proton scattering}
\author{D.A. Fagundes, M.J. Menon, G.L.P. Silva}

\mail{menon@ifi.unicamp.br}          
\institute{
Universidade Estadual de Campinas, 
Instituto de F\'{\i}sica Gleb Wataghin,
13083-859  Campinas, SP, Brazil}
\date{Received:  / Revised version:}
%

\abstract{New developments in empirical analyses of the proton-proton differential cross section
data at high energies are reported.
Making use of an unconstrained model-independent parametrization for the scattering amplitude
and two different fit procedures, all the experimental data 
in the center-of-mass energy interval 19.4 - 62.5 GeV  are quite well described
(optical point and data above the region of Coulomb-nuclear
interference).
The contributions from the real and imaginary parts of the amplitude beyond the
forward direction are discussed and compared with the results from previous analyses
and phenomenological models. Extracted overlap functions (impact parameter space) are outlined
and a critical discussion on model-independent analyses and results are also presented.}

%
\authorrunning{D.A. Fagundes, M.J. Menon, G.L.P. Silva }

\titlerunning{Model-independent data reductions of elastic proton-proton scattering }

\maketitle

\textbf{Contents}

1. Introduction

2. Physical quantities and overlap functions 

3. Improvements in previous analyses

\ \ \ 3.1 Experimental data

\ \ \ 3.2 Unconstrained analytical parametrization

\ \ \ 3.3 Fit procedures

4. Fit results

\ \ \ 4.1 Method 1

\ \ \ 4.2 Method 2

5. Discussion

\ \ \ 5.1 Scattering amplitude

\ \ \ 5.2 Impact picture

\ \ \ 5.3 Conclusion on the fit results

\ \ \ 5.4 Critical remarks

6. Final conclusions and outlooks

\section{Introduction}
\label{sec:1}

Elastic scattering is the simplest kinematic process in had\-ronic collisions
and the new data on proton--proton scattering to be obtained at the CERN-LHC 
by the TOTEM\footnote{TOTal Elastic and diffractive cross section Measurement.}
Collaboration \cite{totem} has brought a renewed interest in the subject. In the theoretical
(dynamical) context, high-energy
elastic hadron scattering, as a soft process, constitutes a topical problem
for QCD: perturbative techniques do not formally apply and nonperturbative
approaches still depend on model assumptions, most of them unjustified
\cite{pred,fiore}. On the other hand, the experimental data available can be well described
by a wide variety of phenomenological models, but with different physical pictures \cite{fiore}
and therefore, an effective theoretical description of these processes, widely accepted, still remains an open problem.
As recently stated,
at this stage, ``empirical 
parametrization, unbiased by any theoretical prejudice can be useful"
\cite{fiore} and that is the point we are interested in here.

We have already treated model-independent analyses of elastic hadron scattering in
a series of works \cite{fms10,fm10,am,sma,acmm,cmm,cm}. The goal is to extract from data empirical
information on what is theoretically unknown, so as to justify some phenomenological inputs,
getting insights for realistic model developments \cite{am} and possible connections with QCD.

The basic approach consists in the introduction of a model-independent
parametrization for the scattering amplitude, fits to the differential cross section data
and empirical extraction of quantities of interest, as for example in an unitarized scheme, the overlap
functions and the eikonal (impact parameter and momentum transfer spaces).
However, that is not a trivial task for at least three main reasons:
(1) the available  data (differential cross sections) cover finite regions
of the momentum transfer and the analytical integrations must be carried out through all
the (physical) momentum transfer space; (2) experimental information on the phase of the
amplitude is available only in the forward direction; (3) in a non-linear fit, the final
result strongly depends on the choice of initial values of the free parameters which are unknown.
As a consequence, it is not possible to reach a unique solution, but only one or more 
\textit{possible solutions}.

The first problem can be treated by selecting only data with the largest values of
momentum transfer available, roughly above $\sim$ 5 GeV$^2$, and that limits the analysis 
to proton--proton ($pp$)
elastic scattering in the energy region 19.4 - 62.5 GeV, as we shall discuss.
The other two problems are crucial, demanding detailed analyses of different solutions,
connected with different parametrization and fit procedures.

We are presently investigating this subject and in this work we discuss new developments,
with better statistical results than those obtained in previous analyses \cite{am,cmm}.
However, we stress that we do not intend to give here the solution, but
display and discuss \emph{what kind of solutions we can arrive with some specific
methodology}. As we shall discuss and explain along the text, only a global comparative
investigation of different methods, parametrizations and solutions, extracting
empirical information on what is common to all cases, can give strong
support for theoretical developments.
We hope that this strategy and the results here presented can contribute with further developments
in the investigation of the so important inverse problems in high-energy elastic hadron scattering.

The work is organized as follows.
In Sect. 2 we recall the main formulas connecting the physical quantities, scattering
amplitude and the overlap functions (impact parameter space).
In Sect. 3 we treat the improvements in previous analyses and  in Sect. 4
we present the fit results. In Sect. 5 we discuss all the results, including comparisons with previous analyses and predictions from phenomenological models,
an outline of the emerging impact picture, our partial conclusions
and some critical comments. The final conclusions and  outlooks are the contents of Sect. 6.

\section{Physical quantities and overlap functions}
\label{sec:2}

\subsection{Physical quantities}

The physical quantities are expressed in terms of the complex elastic scattering amplitude
$A(s,q) = {\rm Re}\ A(s,q) + 
\rm{i}\ {\rm Im}\ A(s,q)$, with
$q^2 = -t$, where $s$ and $t$ are the Mandelstam variables. In this analysis we use 
as input the experimental data only on the  differential cross section,
\begin{eqnarray}
\frac{d\sigma}{dq^2}(s,q) = \pi\ |A(s,q)|^2, 
\end{eqnarray}
the total cross section (Optical Theorem)
\begin{eqnarray}
\sigma_{tot}(s) = 4\pi\ {\rm Im} A(s, q=0), 
\end{eqnarray}
and the $\rho$ parameter
\begin{eqnarray}
\rho(s) = \frac{{\rm Re} A(s,\ q=0)}{{\rm Im} A(s,\ q=0)}.
\end{eqnarray}
From (1) to (3), the optical point is given by
\begin{eqnarray}
\left.\frac{d\sigma}{dq^2} \right|_{q^2=0} = 
\frac{\sigma_{tot}^2 (1 + \rho^2)}{16\pi}. 
\end{eqnarray}

\subsection{Impact parameter space}

The impact parameter representation can be characterized by two basic functions, 
the Profile  $\Gamma(s, b)$ and Eikonal $\chi (s,b)$, where $b$ is the impact parameter. 
In the case of azimuthal symmetry we have \cite{pred}

\begin{eqnarray}
A(s,q) = i \int_{0}^{\infty} bdb J_{0}(qb) \Gamma(s, b),
\end{eqnarray}
where $J_0$ is the zero-order Bessel function and
\begin{eqnarray}
\Gamma(s, b) = 1 - e^{i\,\chi (s,b)}.
\end{eqnarray}

The Unitarity Principle, in the impact parameter space, can be expressed in terms
of the total, elastic and inelastic overlap functions,

\begin{eqnarray}
G_{tot}(s, b) = G_{el}(s, b) + G_{in}(s,b),
\end{eqnarray}
which, in terms of the profile function reads \cite{pred}

\begin{eqnarray}
2\mathrm{Re} \Gamma(s,b) = | \Gamma(s,b) |^2 + G_{in}(s, b),
\end{eqnarray}
where $G_{in}(s, b)$ can be interpreted as the probability
of an inelastic event to take place at $s$ and $b$. From Eqs.
(6) and (8),
\begin{eqnarray}
G_{in}(s, b) = 1 - e^{- 2\,\mathrm{Im}\ \chi (s,b)},
\end{eqnarray}
where $\mathrm{Im}\ \chi (s,b) \geq 0$
and in the black disk limit $G_{in} \rightarrow 1$.

As commented in our introduction, the model-independ\-ent approach consists, essentially, 
in an empirical pa\-ram\-e\-tri\-za\-tion for the amplitude $A(s,q)$, fits to the $d\sigma/dq^2$
data and the extraction of the overlap and eikonal functions \cite{fms10,fm10,am,sma,cmm,cm}.

\section{Improvements in previous analyses}
\label{sec:3}

In this Section we discuss the improvements introduced in the
previous analyses by \'Avila--Menon \cite{am} and Carvalho--Martini--Menon \cite{cmm}, 
which concern three aspects:
compilation and normalization of data at 19.4 GeV (Sect. 3.1), a novel unconstrained
empirical parametrization for the scattering amplitude (Sect. 3.2) and new fit procedures
and results
(Sect. 3.3).

\subsection{Experimental data}

\subsubsection{General aspects}

As already noted, the need of experimental information on the differential
cross sections at intermediate and large values of the momentum transfer
(up to and above $\sim$ 5 GeV$^2$) limits the analysis to $pp$ scattering
at the highest energies with available data. Therefore, 
as in \cite{am}, we treat here
6 sets of experimental data at the energies $\sqrt s$ = 19.4 
(from the Fermilab and CERN-SPS), 23.5,
30.7, 44.7, 52.8, 62.5 GeV (from the CERN-ISR).
Each set includes the optical point, Eq. (4), and all the data above 
the region of Coulomb-nuclear interference, namely $q^2\ >\ $ 0.01 GeV$^2$.

Since the differential cross section data at large momentum
transfer, $q^2\ >\ $ 3.5 GeV$^2$, do not depend on the energy at the ISR region,
following \cite{am,cmm},
we have included in each of the above five sets (ISR), the data 
at $\sqrt s$ = 27.4 GeV (Fermilab), which cover the region 5.5 $\leq$ q$^2$ $\leq$ 
14.2 GeV$^2$ (see \cite{am} for a detailed discussion on this respect
and a complete list of references).

The data from the ISR
have been compiled, analyzed and normalized by Amaldi and Schubert \cite{as}
and the ensemble of elastic $pp$ data is the same used and quoted in \cite{am}, except for
the compilation and normalization of the data at 19.4 GeV, as discussed in what follows.

\subsubsection{Data at 19.4 GeV}

Previous analyses have shown that fit results at 19.4 GeV present some
differences if compared with those obtained at the ISR region.
These differences concern both the goodness of the fit and some extracted
quantities. For example, evaluating the reduced chi-square, $\chi^2$ per degrees of
freedom (DOF), and
taking into account only the statistical errors,
typical results indicate $\chi^2$/DOF $\sim$ 2.76 at 19.4 GeV and an average
1.48 $\pm$ 0.38 with the ISR ensemble \cite{am}. Moreover, as shown in \cite{am},
the extracted eikonal in the momentum transfer space presents a zero (change
of signal) with the ISR data, but not at 19.4 GeV.
For those reasons, we have looked for  all data available at 19.4 GeV and have
investigated the necessity/possibility of a normalization.

First, concerning the compilation, in addition to the data used in \cite{am},
and following \cite{kfk}, we have included those
obtained by Kuznetsov et al. \cite{kuznetsov} and Schiz et al. \cite{schiz}, in the region of 
small momentum transfer. All the compiled data at this energy are listed in 
Table 1. The optical point, Eq. (4), is the same
used in \cite{am}, namely $77.66 \pm 0.02$ mb\,GeV$^{-2}$, from 
$\sigma_{tot} = 38.98 \pm 0.04$ mb \cite{carrol}
and $\rho = 0.019 \pm 0.016$ \cite{fajardo}.

\begin{table}
\begin{center}
\caption{Compiled differential cross section data at 19.4 GeV,
above the Coulomb-nuclear interference region:
interval in momentum transfer, number of points (N) and references.}
\label{tab:1}
\begin{tabular}{ccc}
\hline
 $q^2$ interval (GeV$^2$) & N & Reference \\
\hline
0.0102 - 0.0315 & 24 & Kuznetsov et al. \cite{kuznetsov} \\
0.021 - 0.66 & 133 & Schiz et al. \cite{schiz} \\
0.075 - 3.25 & 55 & Akerlof et al. \cite{akerlof} \\
0.6125 - 3.90 & 33 & Fidecaro et al. \cite{fidecaro} \\
0.95 - 8.15 & 34 & Rubinstein et al. \cite{rubinstein} \\
5.0 - 11.9 & 34 & Faissler et al. \cite{faissler} \\
\hline
\end{tabular}
\end{center}
\end{table}

In a second step, without taking into account the systematic errors, we have checked the 
consistence (normalization) of each data set with those at the nearby
regions of momentum transfer (Table 1) and with the optical point (given above). In what follows we shall refer
to each data set using the surname of the first author.
Specifically, we have checked the consistence of
the extrapolation of the data by  Kuznetsov with the optical point, the data by  Kuznetsov
with those by Schiz  and so on (Table 1), up to the data by Faissler. Figure 1 illustrates the sets at the
diffraction peak (including the optical point) and in the largest region of momentum transfer. All sets are consistent from the optical point up to that by
Rubinstein. However, as shown in the figure, the last set (Faissler)
lies somewhat above the trend from the Rubinstein data.
In this case we have taken into account the systematic error of
15 $\%$ in the Faissler data \cite{faissler}, scaling down the published results by a factor of 0.85.

\begin{figure}[h!]
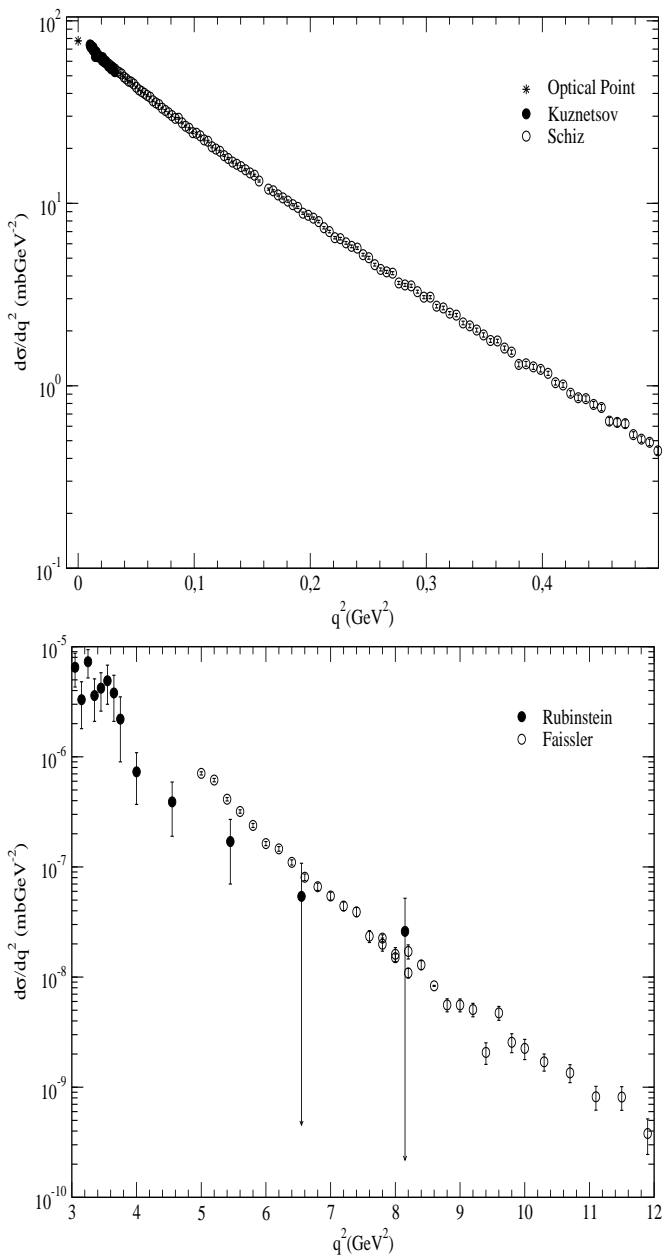

\resizebox{0.48\textwidth}{0.35\textheight}{\includegraphics*{f1a.eps}}
\vspace{0.2cm}
\resizebox{0.48\textwidth}{0.35\textheight}{\includegraphics*{f1b.eps}}
\caption{Differential cross section data at 19.4 GeV at small
(above) and large (below) values of the momentum transfer (see Table 1).}
\label{fig:1} 
\end{figure}

\subsubsection{Data ensemble}

Summing up, in this work the ensemble of elastic $pp$ data is the same used 
and quoted in \cite{am}, except for
the compilation and normalization of the data at 19.4 GeV.
The new compilation at this energy includes
the data by Kuznetsov and Schiz  in the
region of small momentum transfer (0.01--0.66 GeV$^2$) and the normalization
concerns a scale factor of 0.85 in the data by Faissler (largest region of momentum transfer).
With this ensemble the first problem referred to in our introduction
can be partially resolved, since the data cover the region up to
$\sim$ 12 GeV$^2$ at 19.4 GeV and up to $\sim$ 14 GeV$^2$ at the
ISR energy region (23.5--62.5 GeV).

\subsection{Unconstrained analytical parametrization}

In our previous work, we have used a constrained parametri\-zation for
the scattering amplitude, in the sense that
the free parameters in the real
part are also present in the imaginary part \cite{am}: 
\begin{eqnarray}
A(s,q) =  \frac{\rho\, \sigma_{\mathrm{tot}}}{4\pi \sum_{j=1}^{m} \alpha_{j}}
\sum_{j=1}^{m} \alpha_{j} e^{-\beta_{j} q^{2}}
+
\mathrm{i}
\sum_{j=1}^{n} \alpha_{j} e^{-\beta_{j} q^{2}},\quad
\label{eq:11}
\end{eqnarray}
with
\begin{equation}
\sum_{j=1}^{n} \alpha_j = \frac{\sigma_{\mathrm{tot}}}{4 \pi},
\label{eq:12}
\end{equation}
where $m < n$ and $\rho$, $\sigma_{\mathrm{tot}}$ are the experimental values
at each energy. As shown in \cite{am}, with this constraint the contribution
from the real part of the amplitude presents a zero (change of signal) at small
values of the momentum transfer, as predicted in a theorem by A. Martin for
even-signature amplitudes \cite{martin}.

In order to consider
a more general and unconstrained form, we have introduced the
following model-independent parametrization \cite{sma}:
\begin{eqnarray}
&A&(s,q) = \left\{ \left[ {\rho \,\sigma_{tot} \over 4 \pi} -
\sum_{i=2}^{m} a_i \right] e^{- b_1 q^{2}} +
\sum_{i=2}^{m} a_i e^{- b_i q^{2}} \right\}\ \ \nonumber \\
&+&\mathrm{i}\left\{
\left[ {\sigma_{tot} \over 4 \pi} -
\sum_{j=2}^{n} c_j \right] e^{- d_1 q^{2}} +
\sum_{j=2}^{n} c_j e^{- d_j q^{2}} \right\},
\end{eqnarray}
where $a_i$, $b_i$ (real part) and $c_j$, $d_j$ (imaginary part) are now independent
free parameters. In the above formula,
the parameters $a_1$ (real part) and $c_1$ (imaginary part)
have been eliminated through Eqs. (2) and (3). As before, 
$\rho$ and $\sigma_{\mathrm{tot}}$ are experimental inputs at each energy analyzed.

\subsection{Fit procedures}

Besides using the above unconstrained analytical function,
our primary concern here is to develop fit procedures unbiased by any
phenomenological (theoretical) assumption. In this section we discuss
some critical aspects involved and after that, the methodology is developed.

\subsubsection{General aspects}

\begin{quote}
``Fitting nonlinear functions to data samples sometimes seems to be more an
art than a science." \cite{bevington}
\end{quote}

Our purpose is to use the unconstrained parametrization  (12) to fit
the differential cross section data, Eq. (1), at the six energies investigated.
Since each data set comprises $\sim$ 150--300 points (covering 10 decades of
experimental data) and the fits are non-linear in the exponential parameters, we have 
no unique solution and we expect more than 10 free fit parameters at each energy.

For comparison among different fit results we shall use the
$\chi^2$ test for goodness of the fit. Since this test is based on the assumption
of a Gaussian distribution of the data uncertainties \cite{bevington,pugh}, we do not consider systematic errors
of the data points,
but only the statistical errors (except for the 0.85 normalization in the Faissler data
at 19.4 GeV). As a consequence, typical values of the $\chi^2$ per degrees of freedom (DOF), for 100--200
DOF, are not so close to 1, as expected in terms of confidence intervals. However, even in this case, it represents a good test for comparative
discussion on different results obtained with the same criterion.
Anyway, if the systematic errors are included (for example in quadrature with the statistic errors),
one obtains $\chi^2$/DOF $\sim$ 1, which seems to us unjustified on statistical grounds.

Let us now discuss the other two main problems referred to in our introduction,
namely the non-linearity of the fit and the contributions from the real and imaginary parts of the
amplitude, which are unknown beyond the forward direction.

The non-linearity  implies in a multimodal Chi-square function of the
2(m + n -1) free parameters, that is, a hypersurface with multiple local minima and
no direct access to a global minimum. As a consequence, any fit demands the choice of initial values
of the parameters, which are also unknown and may lead to a local minimum without physical meaning
and/or multiple solutions. 

Although parametrization (12) brings enclosed the full experimental information on the forward amplitude
at each energy (namely Re $A(s,0)$ and Im $A(s,0)$ from $\sigma_{tot}$ and $\rho$ data), the phase
of the amplitude beyond this region is experimentally unknown. Therefore, looking here for a completely unbiased fit,
we shall not consider any constraint in initial values of the parameters associated with both
the real and imaginary parts of
the amplitude, nor put any limit in the number of parameters involved. 
Moreover, we are not interested to extract any energy dependence of the free parameter.
The main point is to obtain a good fit on statistical grounds, at least better than those 
obtained in previous analyses \cite{am,cmm}.

The non-linear fits have been
performed using the code CERN-Minuit \cite{minuit}, through successive runs of the MIGRAD minimizer and
with the confidence level for the
uncertainties in the free parameter fixed at 70 \% (variable up parameter depending on the number 
of degrees of freedom). The error-matrix provides the variances and covariances associated with
each free parameter. This information is used in the evaluation of uncertainty regions in the
differential cross section and all the extracted quantities, through standard error propagation
procedures \cite{bevington}.

\subsubsection{Methodology for initial values}

With all the above mentioned aspects in mind, we have considered two independent procedures
for the choice of the initial values of the parameters, which we shall denote Methods 1 and 2.

\noindent
\textit{$\bullet$ Method 1}

In order to develop a first test on the response of our unconstrained parametrization to
different initial values of the  parameters, we based our choices on results already
obtained in two other analysis, but excluding any constraint or assumption involved,
as explained in what follows.

The first one concerns the results by \'Avila and Menon \cite{am}, obtained with the constrained
parametrization (11). For each data set we have used the values of the free parameter
(Table 5 in \cite{am}) as initial values for the fits with our unconstrained parametrization
(12), that is, the parameters in the real and imaginary parts of the amplitude
being now completely independent.

The second one is related to a model-independent analysis by Fran\c ca and Hama,
in the same interval of energy \cite{fh}. The original parametrization is purely
imaginary and the fits indicate a second dip in the region of large momentum transfer.
In our procedure, the values of the free parameters (Table 1 in \cite{fh}) have been
used as initial values in the imaginary part of our unconstrained amplitude. After a first run
we have added one, two, three exponential terms in the real part until the
fit reach the best $\chi^2$/DOF (closest to 1), before starting to increase.

We have also attempted analogous procedure with the parametrization by
Amaldi and Schubert \cite{as}, but that form cannot describe the data
at large values of the momentum transfer (data at 27.4 GeV) and the quality of the fits
was also very poor.

Despite the differences in the structure of the two original parametrization and assumed conditions,
the final results with the unconstrained parametrization and the above procedures were very similar
($\chi^2$/DOF and the contributions from the real and imaginary parts of the amplitudes).
Therefore, we have selected for each energy the best statistical result ($\chi^2$/DOF closest to 1)
independently of the variant considered.
The results are presented and commented in Sects. 4 and 5.

\noindent
\textit{$\bullet$ Method 2}

The second method is completely independent of the first one and has been developed in two steps.

In the first step we have considered only the set at 52.8 GeV, since it represents the best statistical information available (original
data covering the region up to $\sim$ 9.8 GeV$^2$ and with the addition of the
data at 27.4 GeV, up to $\sim$ 14 GeV$^2$, totalling 245 points). The fit has been started
with the data near the forward direction, $q^2 \leq$ 0.1 GeV$^2$ and one exponential term
in the imaginary part of the amplitude. In the logarithmic scale, the slope and the intercept
of the straight line can be directly extracted and used as initial values for the fit.
We have then extended the data up  0.5 GeV$^2$, using the previous values as feed back and testing all possible number of exponential terms in the real and imaginary parts of the amplitude.
Specifically, representing this number of terms as an ordered pair we have tested

\vspace{0.3cm}

\centerline{(real, imaginary) = (0, 1), (1, 1), (1, 2), (2, 1), (2, 2),...,}

\vspace{0.3cm}

\noindent
until the fit result reach the smallest $\chi^2$/DOF, before to increasing again.
After that, with the same procedure, we have extended the data up to 1.5 GeV$^2$ and
then up to 14.2 GeV$^2$, using in each case the previous values as feed back and repeating the
procedure.

In a second step, once obtained the best fit at 52.8 GeV, the values of the parameters
have been used
as initial values for the fits at
the nearby energies, namely from 52.8 to 62.5 GeV, from 52.8 to 44.7 GeV
and then to 30.7, 23.5 and 19.4 GeV.

In all the above procedures, if the uncertainty in a given parameter was of the same order of, or
greater than, its central value, this parameter was excluded and the fit performed again. 
We have also tested different confidence intervals for the uncertainties in the fit
parameters, but the results were not as good as those with the fixed 70 \% CL. Details
on these results can be found in \cite{ge}.

\section{Fit Results}

The final fit results with Methods 1 and 2 are displayed in Tables 2 and 3, respectively. The visual
description of the experimental data are quite similar in both cases and are illustrated in Figure 2,
where it is also shown the uncertainty regions in the differential cross section,
evaluated through standard error propagation from the fit parameters.

\begin{table*}
\begin{center}
\caption{Fit results at each energy analyzed with Method 1: values of the free fit
parameters and corresponding errors (all in GeV$^{-2}$), number of points ($N$),
number of degrees of freedom (DOF) and reduced $\chi^2$.}
\label{tab:2}
\begin{tabular}{cccccccc}
\hline
 & $\sqrt{s}$ (GeV):  & 19.4 & 23.5 & 30.7 & 44.7 &  52.8  & 62.5 \\
\hline
    &    $a_{1}$& -0.03    &  -0.05 &   -0.022    &  -0.035       & -0.043      & -0.022 \\
   &          & $\pm$ 0.13 & $\pm$0.40 & $\pm$ 0.090  & $\pm$ 0.094  & $\pm$ 0.087  & $\pm$ 0.097 \\
   &    $b_{1}$& 0.772 &  0.93       &   0.764      &  0.858       & 0.912       & 0.774 \\
Re$\,A$ &   & $\pm$ 0.034&  $\pm$ 0.10 &  $\pm$ 0.046 &  $\pm$ 0.050 & $\pm$ 0.034 & $\pm$ 0.047 \\
   &     $a_{2}$& 0.1825 &    0.212   & 0.3664      &      0.5646     & 0.7232     & 0.8631 \\
   &   &  $\pm$ 0.0055 & $\pm$ 0.023 & $\pm$ 0.0047 & $\pm$ 0.0063 & $\pm$ 0.0053 & $\pm$ 0.0046 \\
   &   $b_{2}$ & 1.760&  2.01      &   2.644     & 3.12       & 3.249       & 4.16 \\
   &           & $\pm$ 0.061 & $\pm$ 0.18 & $\pm$ 0.092  & $\pm$ 0.13 & $\pm$ 0.099 & $\pm$ 0.37 \\
\hline 
     &  $c_{1}$  & -34.08 &  -32  &  -157.30  &    1244   & -128.8    & -140.1 \\
     &       &  $\pm$ 0.43 & $\pm$ 16 & $\pm$ 0.61 & $\pm$ 22  & $\pm$ 1.0 & $\pm$ 2.7 \\
     &  $d_{1}$& 2.5259 &  2.63 &   2.33972     & 10.2433       & 2.77889    & 2.7678 \\
     & &  $\pm$ 0.0024& $\pm$ 0.15 & $\pm$ 0.00028 & $\pm$ 0.0042 & $\pm$ 0.00087 & $\pm$ 0.0019 \\
     &  $c_{2}$ & 392.00 &  398     &    3.802    &   1460       &    2.098    & 1.87 \\
     &         & $\pm$ 0.31 & $\pm$ 11 & $\pm$ 0.054 & $\pm$  10   & $\pm$ 0.081 & $\pm$ 0.19 \\
     &   $d_{2}$&2.88166 & 2.897   &    8.51    &  9.7419   &   13.25    & 14.4 \\
     &       &  $\pm$ 0.00033 & $\pm$ 0.046 & $\pm$ 0.18 & $\pm$ 0.0034 & $\pm$ 0.54 & $\pm$ 1.3 \\
     &        $c_{3}$& 22.48 & 9.4 & 291.24     &  -4.613   & -4.613   & 226.3\\
     &  &  $\pm$ 0.13  &   $\pm$ 3.1      & $\pm$ 0.070 & $\pm$ 0.070 & $\pm$ 0.94 & $\pm$ 2.6 \\
     &  $d_{3}$ & 4.023 & 4.68    & 2.43478       & 2.283         & 2.89792       & 2.8746\\
 Im$\,A$  &    &  $\pm$ 0.013   & $\pm$ 0.74 & $\pm$ 0.00024 & $\pm$ 0.012   & $\pm$ 0.00051 & $\pm$ 0.0012 \\
     &  $c_{4}$& -373.86 & -369  &    -219.53    &  -2697    & -135.19    & -118.34 \\
     &     &  $\pm$ 0.27 & $\pm$ 11   & $\pm$ 0.19    & $\pm$ 20   & $\pm$ 0.28 & $\pm$ 0.77 \\
     &  $d_{4}$ & 2.94863 &  2.931    &    2.68382 &    9.9791   &       3.2963     &  3.2259 \\
     &      &  $\pm$ 0.00043  & $\pm$ 0.045 & $\pm$ 0.00035 & $\pm$ 0.0020 & $\pm$ 0.0015 & $\pm$ 0.0043 \\
     &$c_{5}$& 1.425  & 1.48 & 89.99      & $1.57\times 10^{-3}$   &    56.30   & 39.16 \\
     & & $\pm$ 0.075 & $\pm$ 0.98  & $\pm$ 0.13 & $\pm 0.84\times 10^{-3}$& $\pm$ 0.18 & $\pm$ 0.44 \\
     &  $d_{5}$& 11.35 &  12.7    &   2.9567    &    0.399    &   3.8306    &  3.942 \\
     &      &  $\pm$ 0.56  & $\pm$ 4.7  & $\pm$ 0.0013 & $\pm$ 0.052 & $\pm$ 0.0052 & $\pm$ 0.020 \\
     &  $c_{6}$ &$ 2.98\times 10^{-3}$ & $1.7\times 10^{-3}$ & $1.15\times 10^{-3}$ & 5.702 &
 $1.79\times 10^{-3}$  & $-1.22\times 10^{-3}$\\
     &   & $\pm$ $0.59\times 10^{-3}$ & $\pm 1.2\times 10^{-3}$  & $\pm 0.57\times 10^{-3}$ & $\pm$
 0.073 & $\pm 0.63\times 10^{-3}$ & $\pm 0.72\times 10^{-3}$\\
     &  $d_{6}$& 0.446 & 0.405    & 0.373      & 2.435       & 0.410       & 0.377 \\
     &  & $\pm$ 0.021 & $\pm$ 0.073  & $\pm$ 0.046 & $\pm$ 0.013 & $\pm$ 0.039 & $\pm$ 0.056 \\
\hline
& N     &  314 & 173   & 212 & 247   & 245  & 164 \\
& DOF            & 300  & 159   & 198 & 233   & 231  & 150 \\
& $\chi^{2}$/DOF     & 2.62   &1.09   & 1.09          & 1.99   &1.55    & 1.18  \\
\hline
\end{tabular}
\end{center}
\end{table*}

\begin{table*}
\begin{center}
\caption{Same legend as in Table 2 with Method 2.}
\label{tab:3}
\begin{tabular}{cccccccc}
\hline
 & $\sqrt{s}$ (GeV):  & 19.4 & 23.5 & 30.7 & 44.7 &  52.8  & 62.5 \\
\hline
      & $a_{1}$& 5168    &  5489 &   1730    &  18.6      & 51.7      & 2780 \\
      &       & $\pm$ 34 & $\pm$85 & $\pm$ 210  & $\pm$ 2.7  & $\pm$ 6.2  & $\pm$ 150 \\
      & $b_{1}$& 7.43058 &  7.4862       &   7.1947      &  8.09      & 8.09       & 8.0228 \\
      & & $\pm$ 0.00055&  $\pm$ 0.0013 &  $\pm$ 0.0128 &  $\pm$ 0.24 & $\pm$ 0.13 & $\pm$ 0.0077\\
      &  $a_{2}$& -1238.3 &    -2182   & -1730      &      -18.0    & -34.8     & -1226 \\
Re$\,A$&      &  $\pm$ 93 & $\pm$ 42 & $\pm$ 210 & $\pm$ 2.7 & $\pm$ 3.4 & $\pm$ 80 \\
      & $b_{2}$ & 7.1578&  7.3353      &   7.2188     & 6.25       & 6.604       & 7.787 \\
      &        & $\pm$ 0.0019 & $\pm$ 0.0027 & $\pm$ 0.0031  & $\pm$ 0.11 & $\pm$ 0.072 & $\pm$ 0.014 \\
      & $a_{3}$& -3930 &    -3307   & --      &      --    & -16.2     & -1550 \\
      & &  $\pm$ 33 & $\pm$ 74      & --      & --         & $\pm$ 5.2 & $\pm$ 120 \\
      & $b_{3}$ & 7.51291 &  7.5767 & --      & --       & 9.75       & 8.190 \\
      &        & $\pm$ 0.00068 & $\pm$ 0.0020 & --      & -- & $\pm$ 0.50 & $\pm$ 0.012 \\
\hline 
     & $c_1$&-5200   &  -570  &  -1520   &  -148    & -80      & -70     \\
     &       & $\pm$ 120& $\pm$ 88 & $\pm$ 170 & $\pm$ 19 & $\pm$ 10 & $\pm$ 14 \\
     & $d_{1}$& 3.99471 &  4.8389 &   5.0777     & 4.968       & 4.909       & 5.024 \\
     &    &$\pm$0.000072 & $\pm$ 0.0037 & $\pm$ 0.0012 & $\pm$ 0.011 & $\pm$ 0.019 & $\pm$ 0.034 \\
     & $c_{2}$&5200 &  579     &  1530    &   157     &    90    & 79  \\
     &       &$\pm$120 & $\pm$ 88 & $\pm$ 170 & $\pm$ 19 & $\pm$ 10 & $\pm$ 14 \\
     &$d_{2}$&3.99892 & 4.8645   &    5.0891    &    5.065    &   5.077     & 5.204 \\
     &   & $\pm$ 0.00013& $\pm$ 0.0040 & $\pm$ 0.0014 & $\pm$ 0.012 & $\pm$ 0.019 & $\pm$ 0.031 \\
     &$c_{3}$ & 8.42& 0.0536 &   0.0525    &    0.0536    & 0.0574   & 0.0467   \\
Im$\,A$ &    &  $\pm$ 0.33 &$\pm$ 0.0051  & $\pm$ 0.0043 & $\pm$ 0.0050 & $\pm$ 0.0043 & $\pm$ 0.0052  \\
      &$d_{3}$& 3.117 & 1.026    & 1.082       & 1.062       & 1.055        & 1.008  \\
      &  & $\pm$ 0.015& $\pm$ 0.056 & $\pm$ 0.056 & $\pm$ 0.056 & $\pm$ 0.038 & $\pm$ 0.057 \\
  & $c_{4}$ &$1.35\times 10^{-2}$ & $1.41\times 10^{-3}$ & $2.17\times 10^{-3}$ &  $1.85\times 10^{-3}$
 &  $1.37\times 10^{-3}$ & $1.41\times 10^{-3}$ \\
 & & $\pm 0.075\times 10^{-2}$ &$\pm 0.66\times 10^{-3}$ & $\pm 0.79\times 10^{-3}$ & $\pm 0.77\times
 10^{-3}$ & $\pm 0.51\times 10^{-3}$ & $\pm 0.64\times 10^{-3}$ \\
  &    $d_{4}$&0.6027 &  0.392     &   0.431    &    0.416    &   0.388     &  0.392 \\
  &     & $\pm$ 0.0074 & $\pm$ 0.046 & $\pm$ 0.038 & $\pm$ 0.042 & $\pm$ 0.038 & $\pm$ 0.044 \\
\hline
 & N     &  314 & 173   & 212 & 247   & 245  & 164\\
 & DOF            &   302   &   161    &  202   &  237     &  233    &  152   \\
 & $\chi^{2}/DOF$     & 2.53   &1.77   & 3.79          & 1.87   &1.55    & 1.18  \\
\hline
\end{tabular}
\end{center}
\end{table*}

\begin{figure}[h]
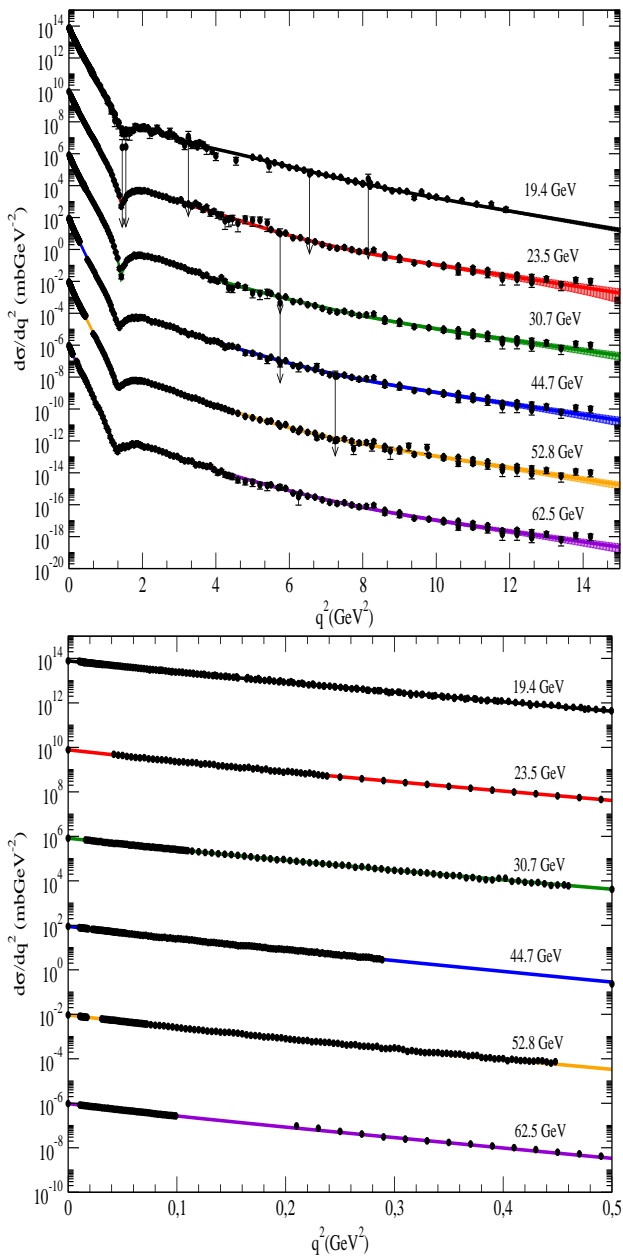

\resizebox{0.45\textwidth}{0.35\textheight}{\includegraphics*{f2a.eps}}
\resizebox{0.45\textwidth}{0.35\textheight}{\includegraphics*{f2b.eps}}
\caption{Fit results and uncertainty regions from error propagation:
all $q^2$ region (above) and diffraction peak (below).
Curves and data have been multiplied by factors of
$10^{\pm 4}$.}
\label{fig:2} 
\end{figure}

The contributions to the differential cross sections from the real and imaginary parts
of the amplitude,
\begin{eqnarray}
\frac{d\sigma^R}{dq^2} = \pi\ [\mathrm{Re}\, A]^2, 
\qquad
\frac{d\sigma^I}{dq^2} = \pi\ [\mathrm{Im}\, A]^2,
\end{eqnarray}
are shown in Figures 3 (Method 1) and 4 (Method 2), where
we display only the uncertainty regions and the experimental data.
In the following we are going to discuss all these results.

\begin{figure*}[h!]
\resizebox{0.5\textwidth}{0.35\textheight}{\includegraphics*{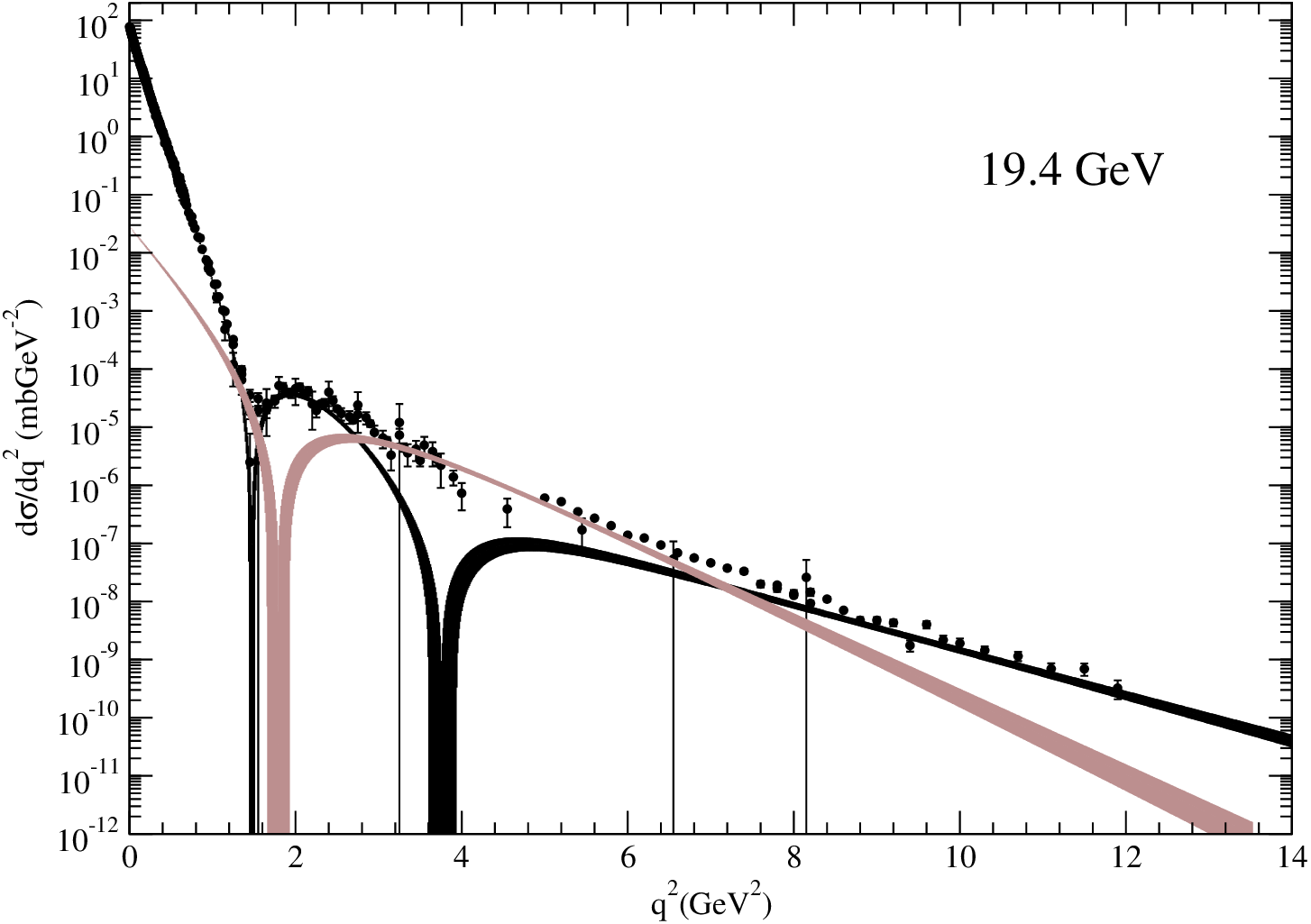}}
\resizebox{0.5\textwidth}{0.35\textheight}{\includegraphics*{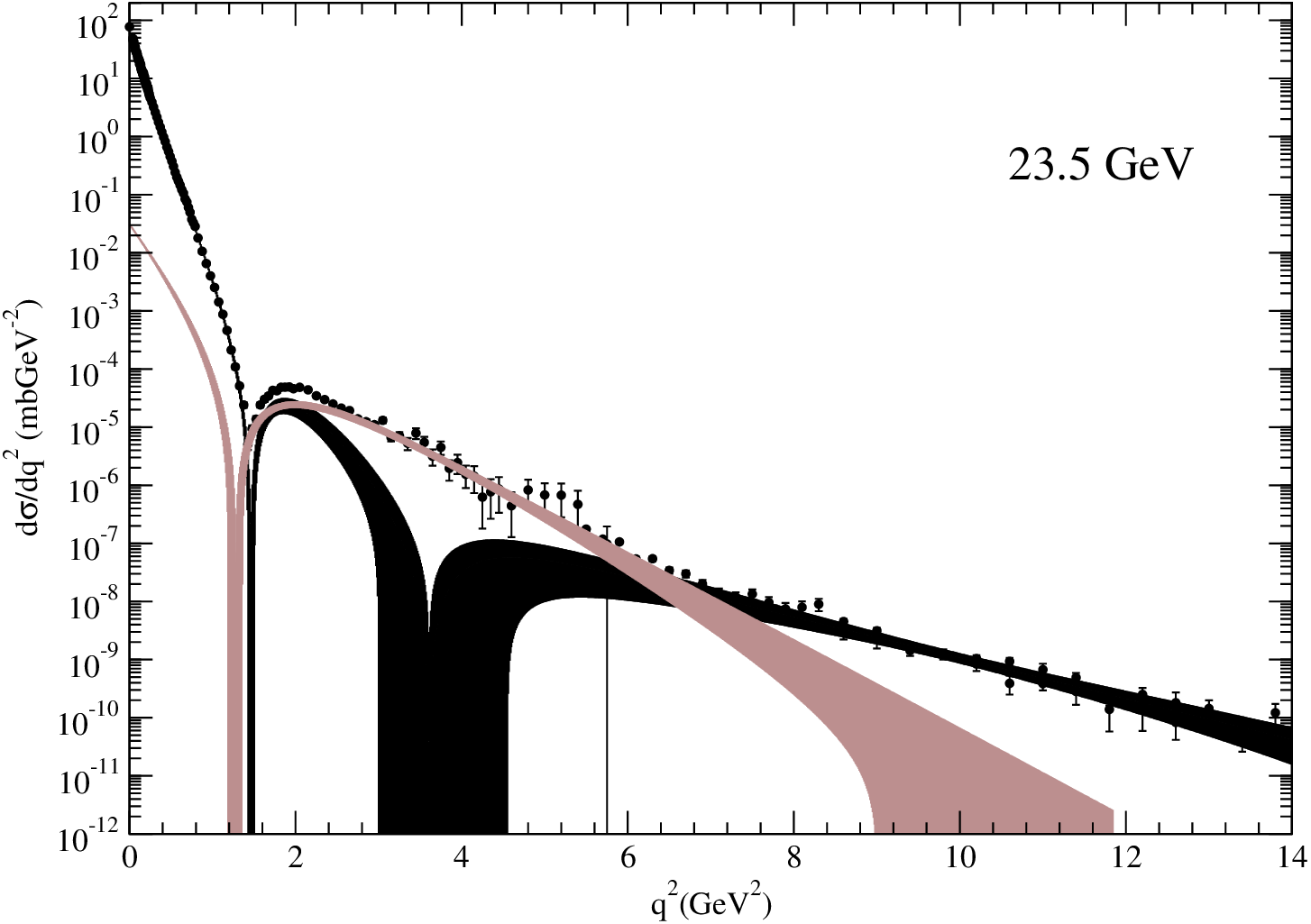}}
\resizebox{0.5\textwidth}{0.35\textheight}{\includegraphics*{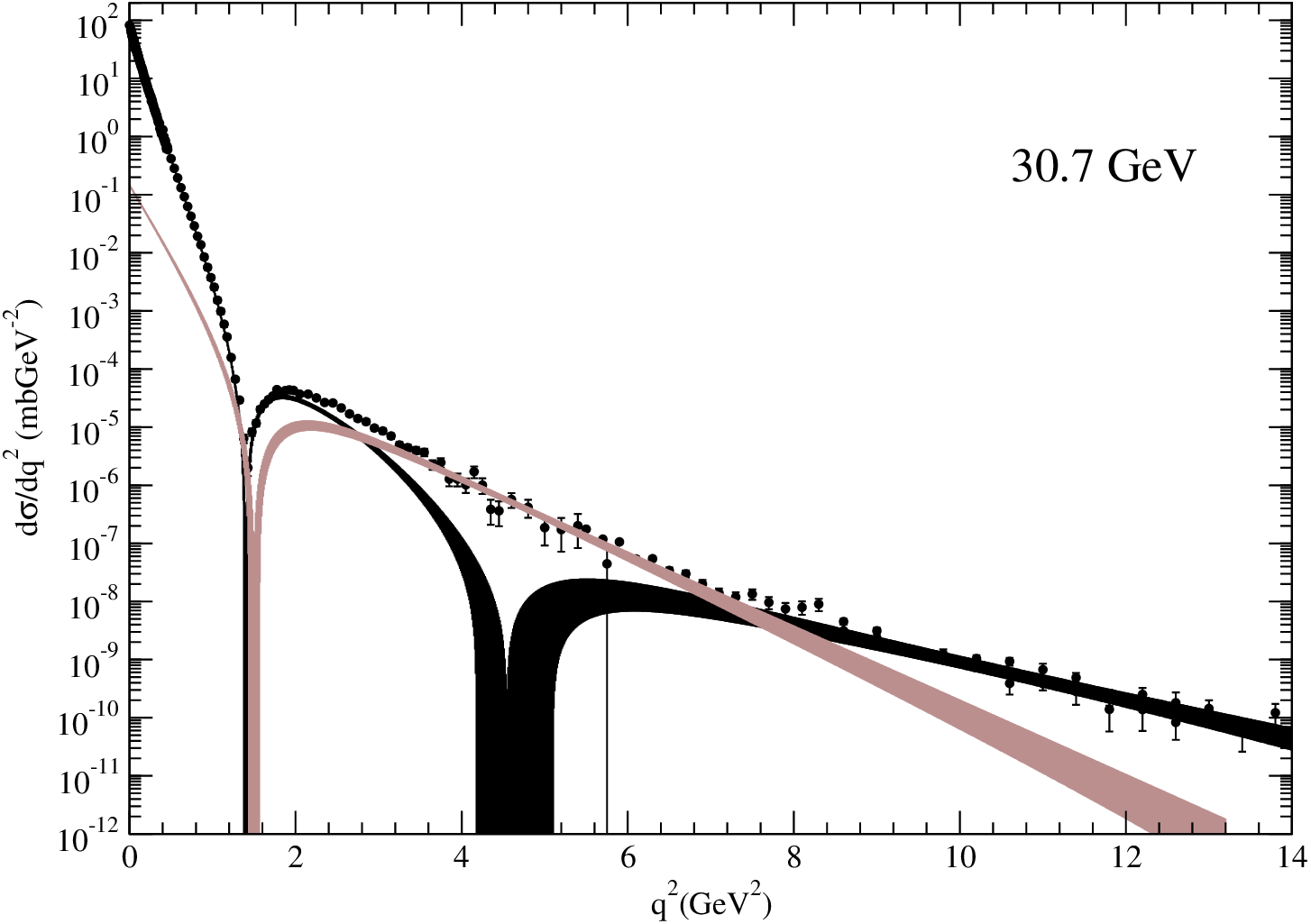}}
\resizebox{0.5\textwidth}{0.35\textheight}{\includegraphics*{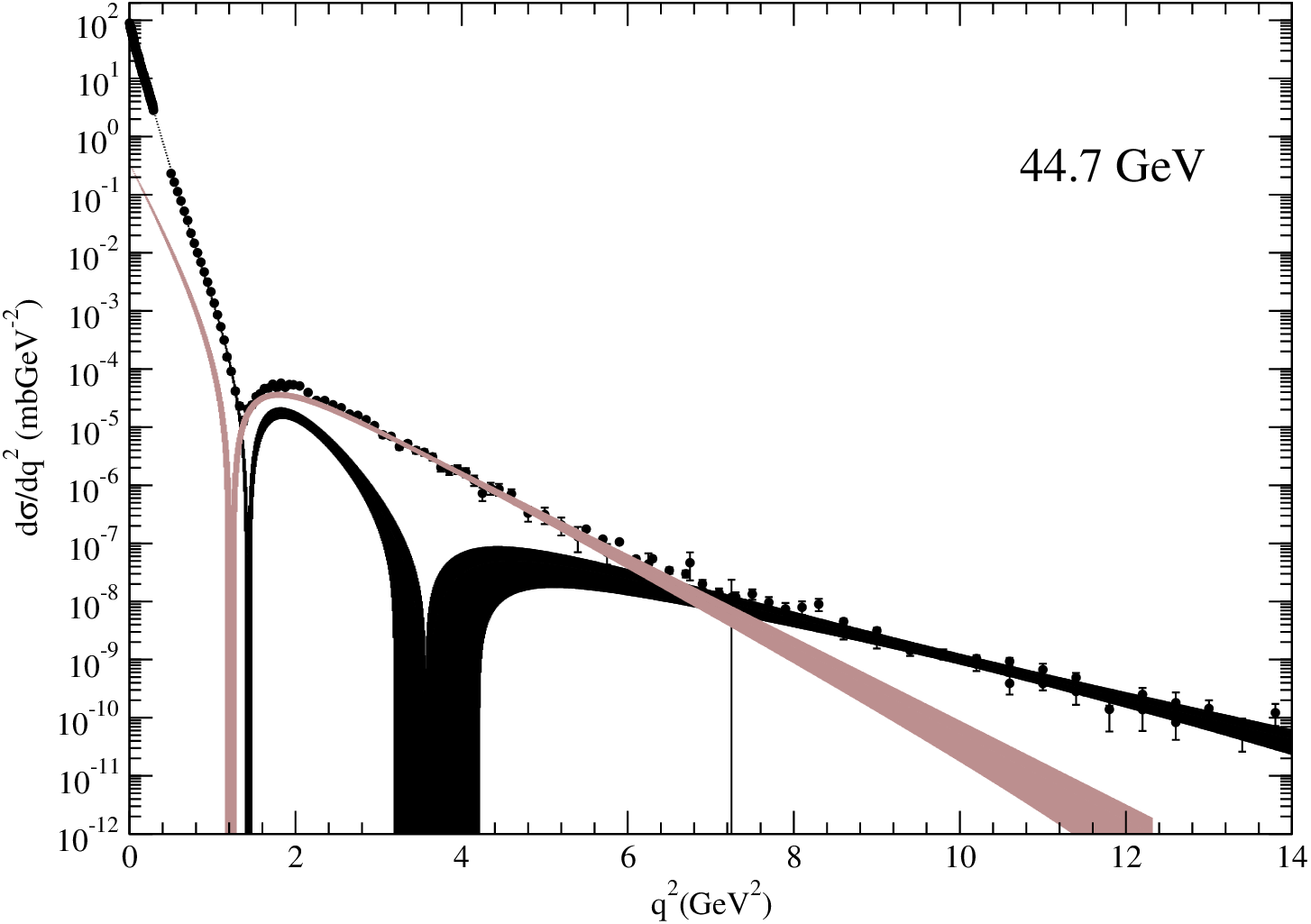}}
\resizebox{0.5\textwidth}{0.35\textheight}{\includegraphics*{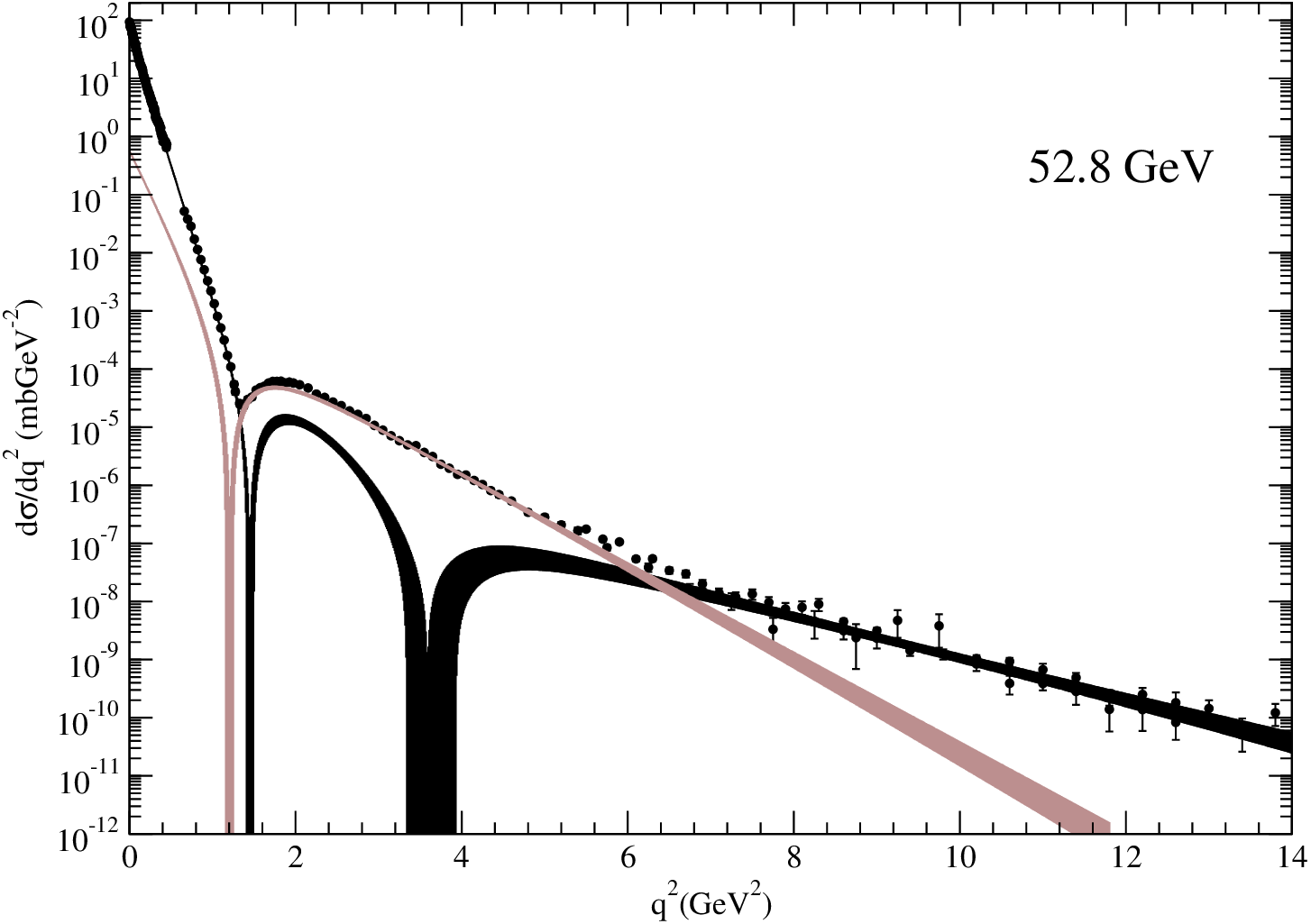}}
\resizebox{0.5\textwidth}{0.35\textheight}{\includegraphics*{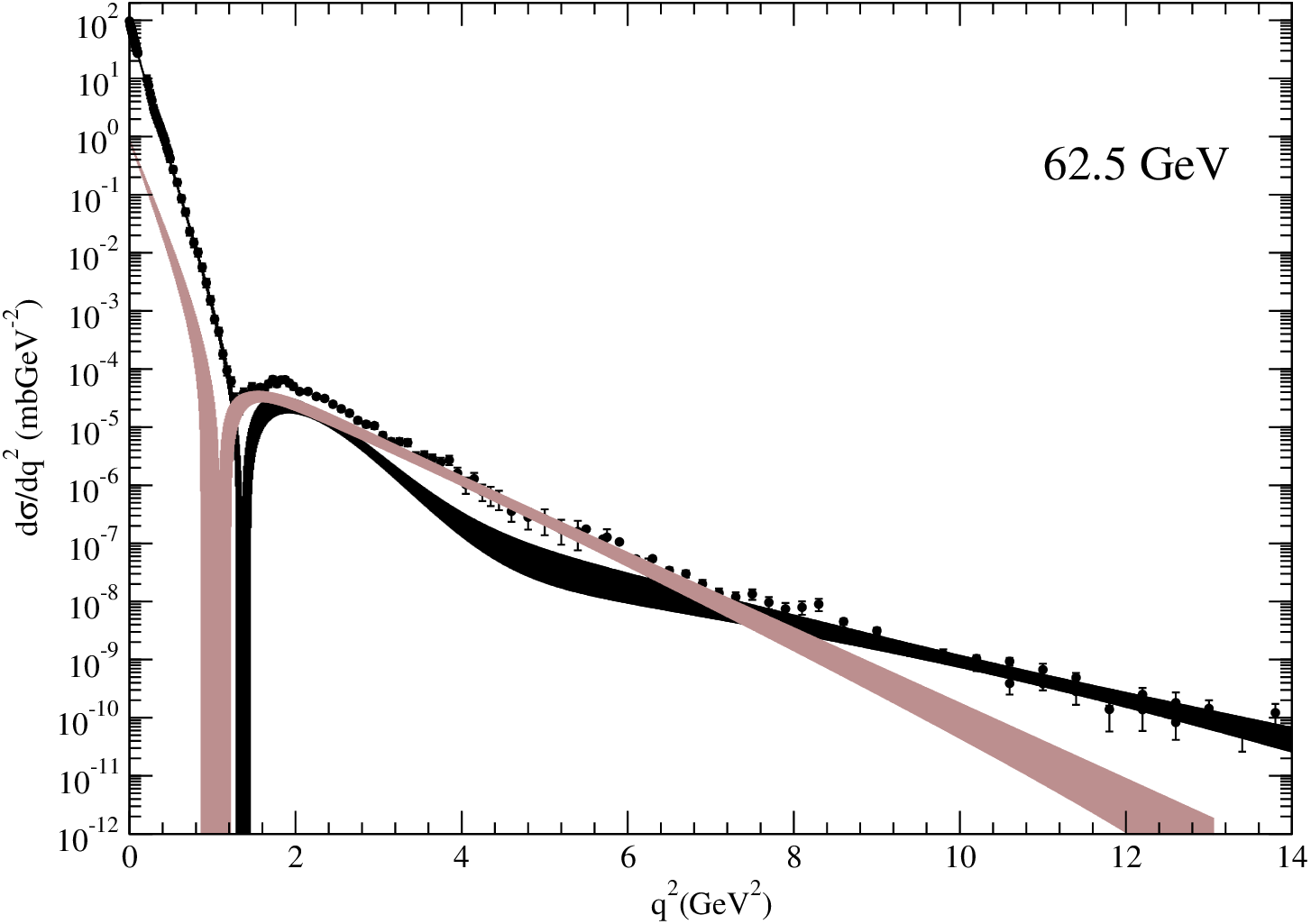}}
\caption{Method 1: uncertainty regions for the contributions to the differential 
cross sections from Re$\,A(s,q)$ (grey) and Im$\,A(s,q)$ (black) and the experimental data.}
\label{fig:3} 
\end{figure*}

\begin{figure*}[h!]
\resizebox{0.5\textwidth}{0.35\textheight}{\includegraphics*{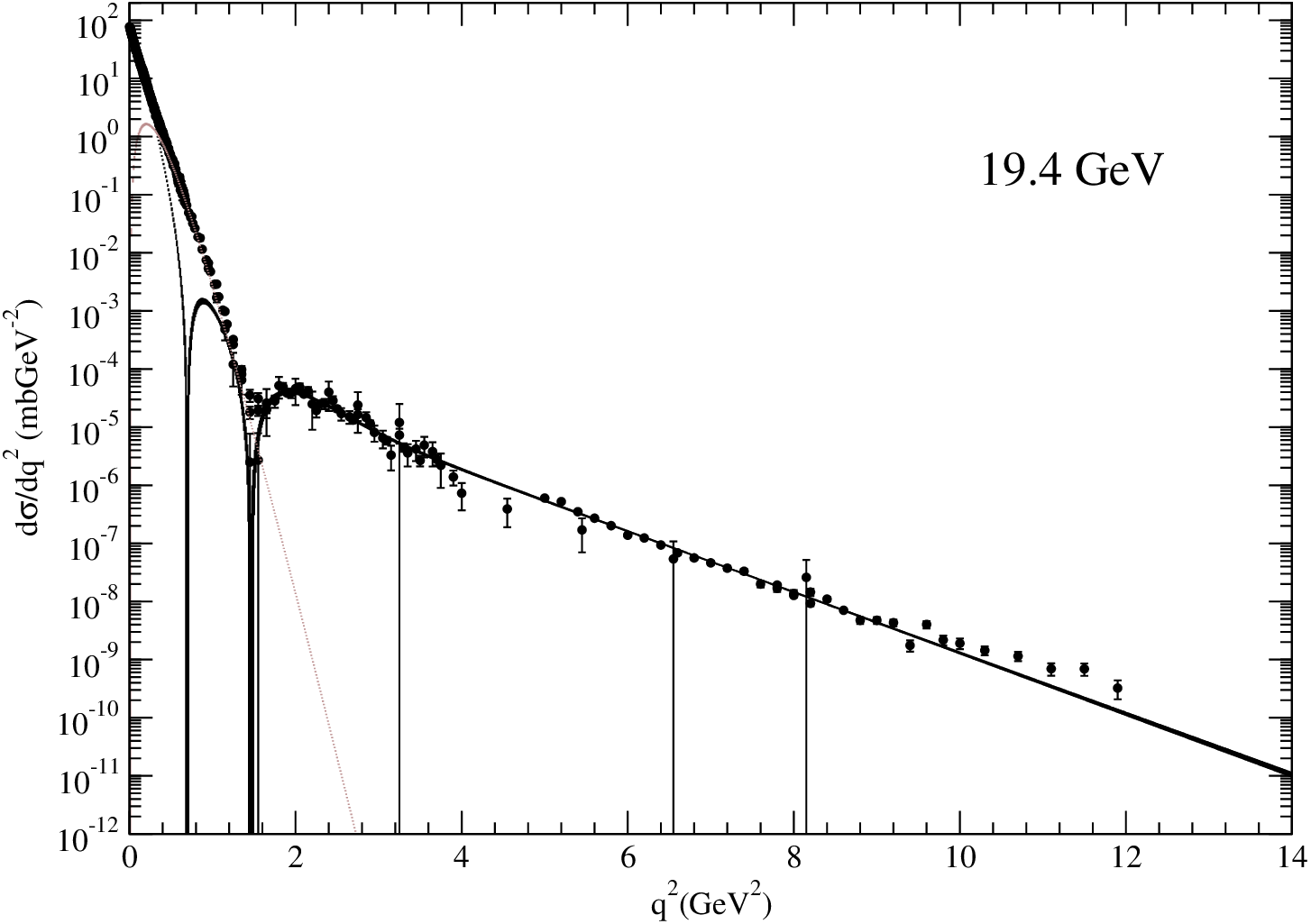}}
\resizebox{0.5\textwidth}{0.35\textheight}{\includegraphics*{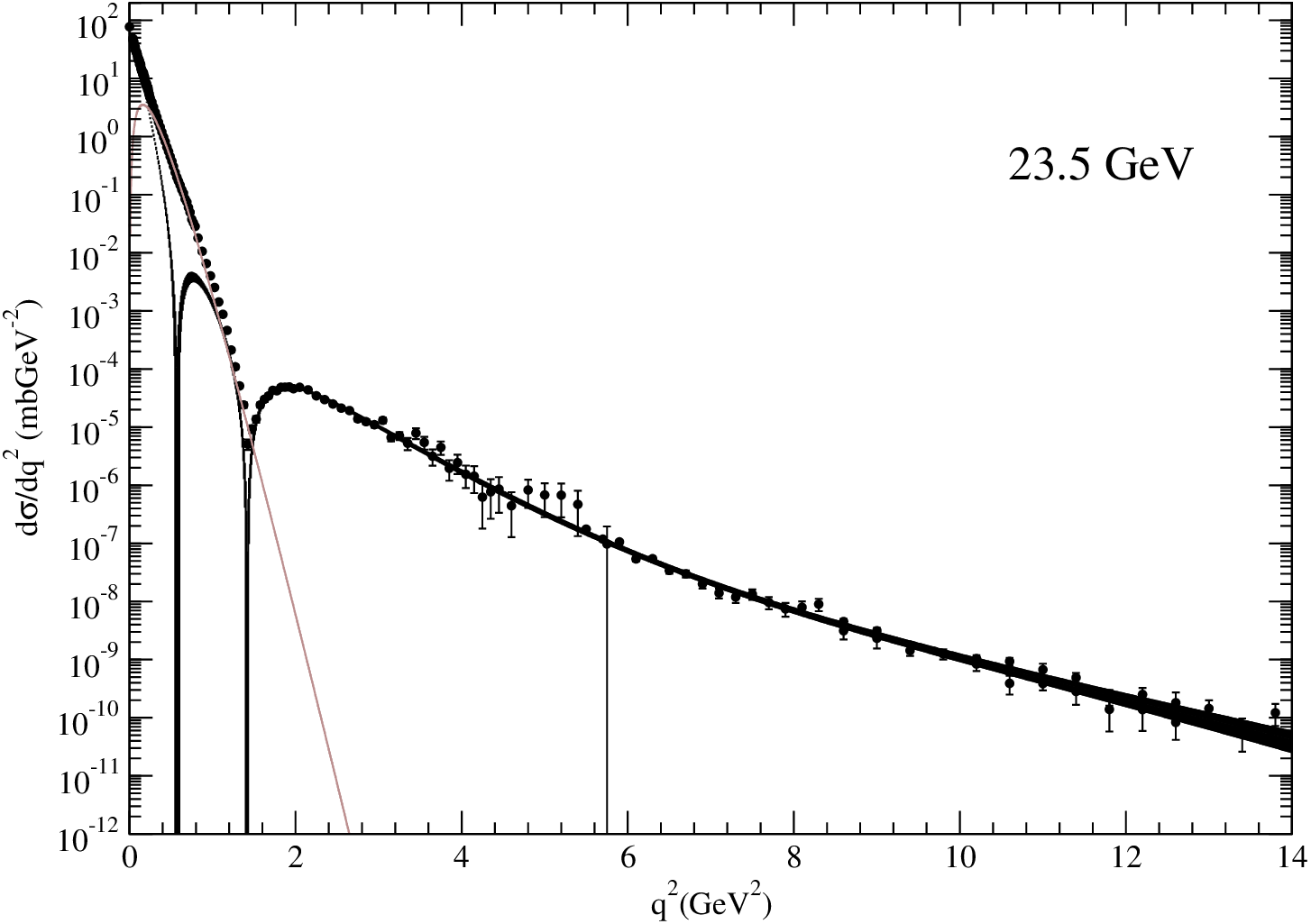}}
\resizebox{0.5\textwidth}{0.35\textheight}{\includegraphics*{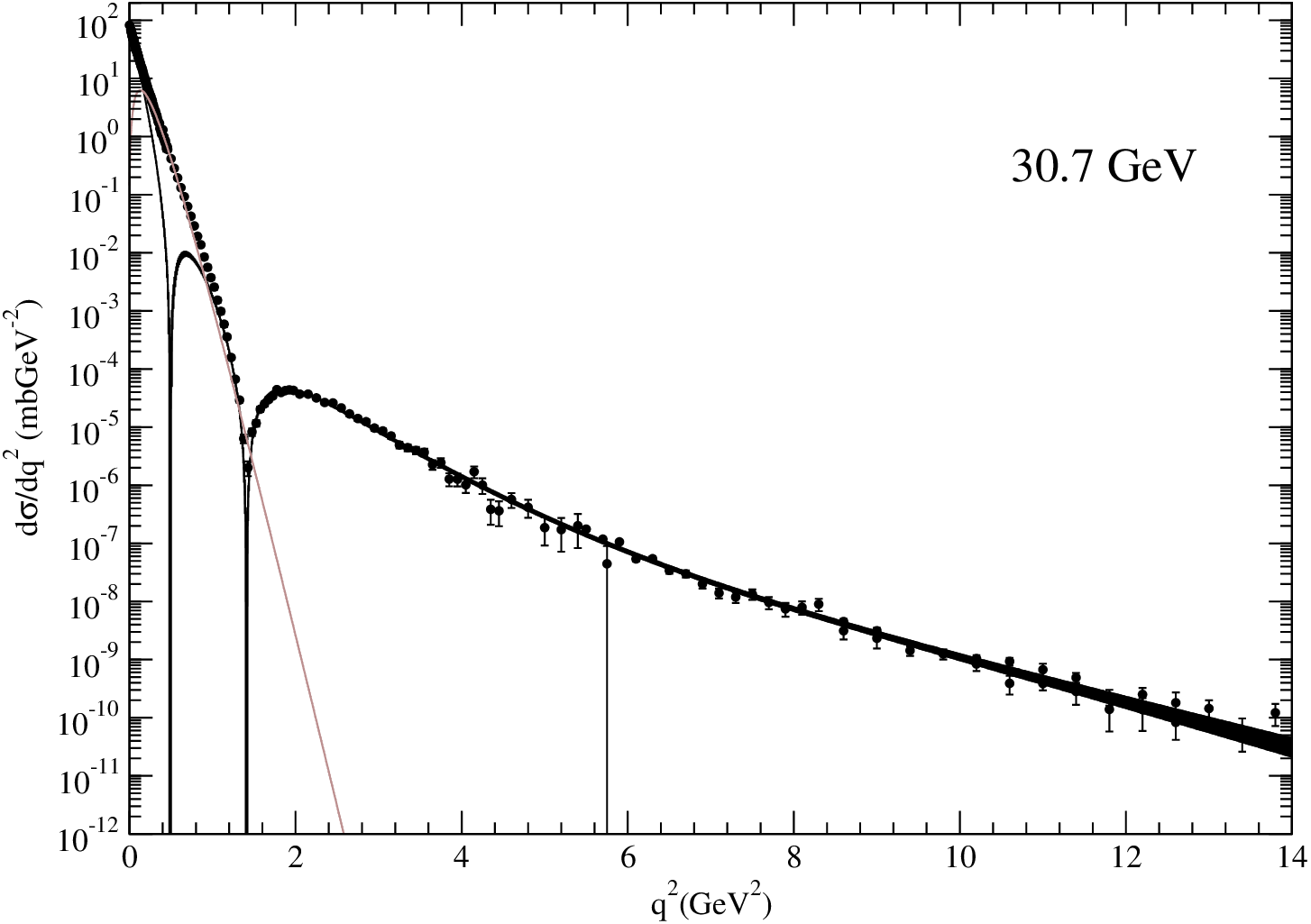}}
\resizebox{0.5\textwidth}{0.35\textheight}{\includegraphics*{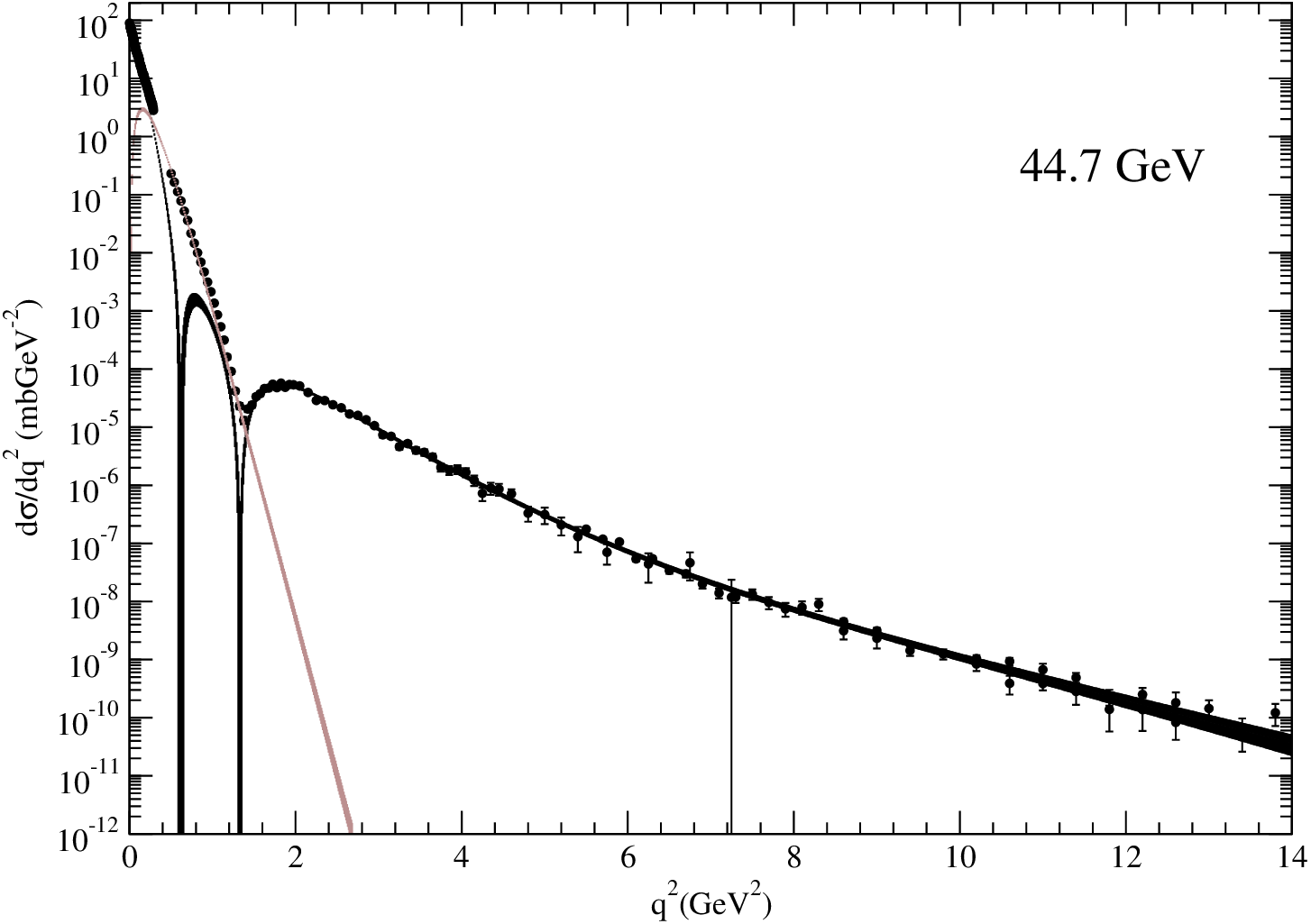}}
\resizebox{0.5\textwidth}{0.35\textheight}{\includegraphics*{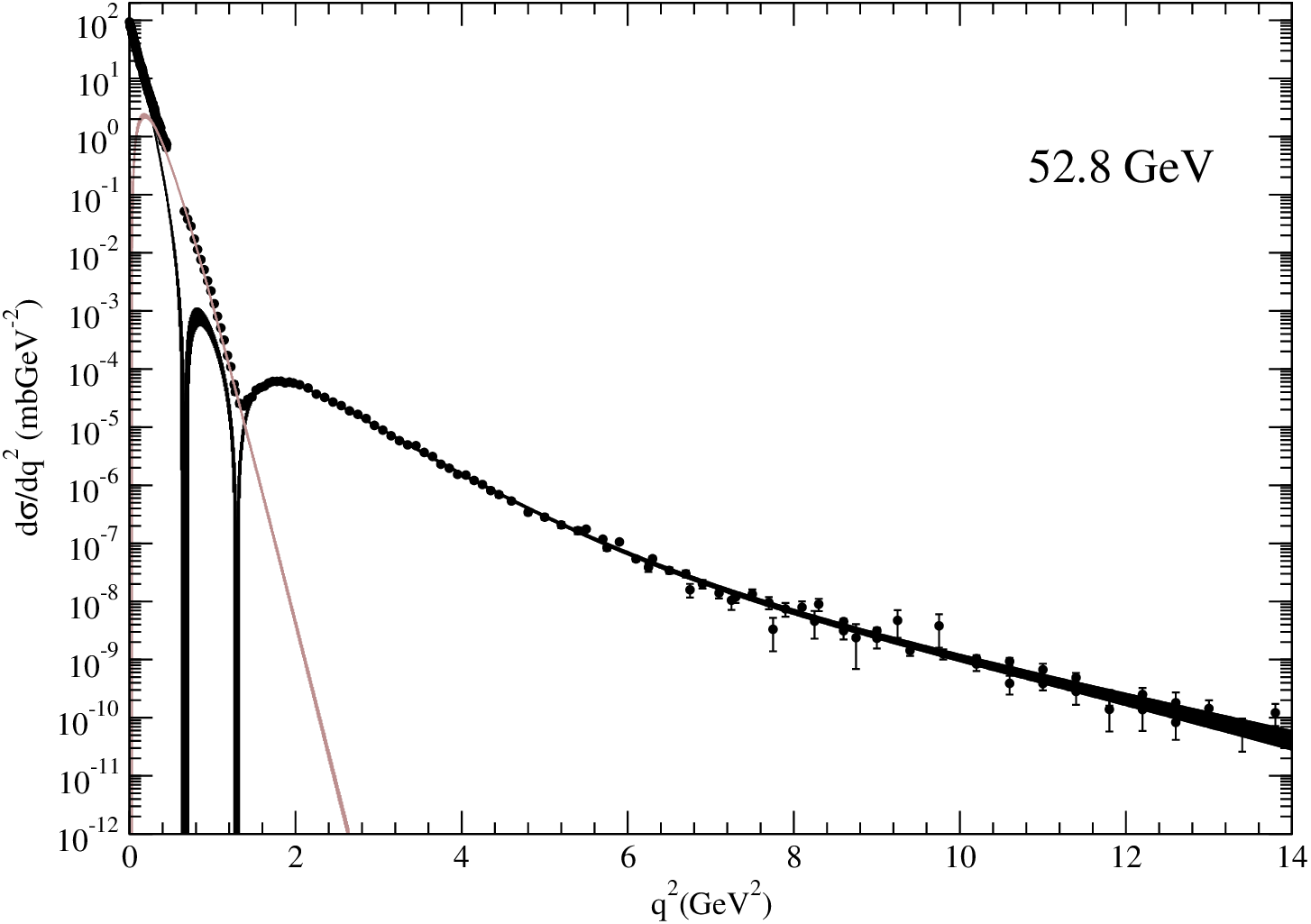}}
\resizebox{0.5\textwidth}{0.35\textheight}{\includegraphics*{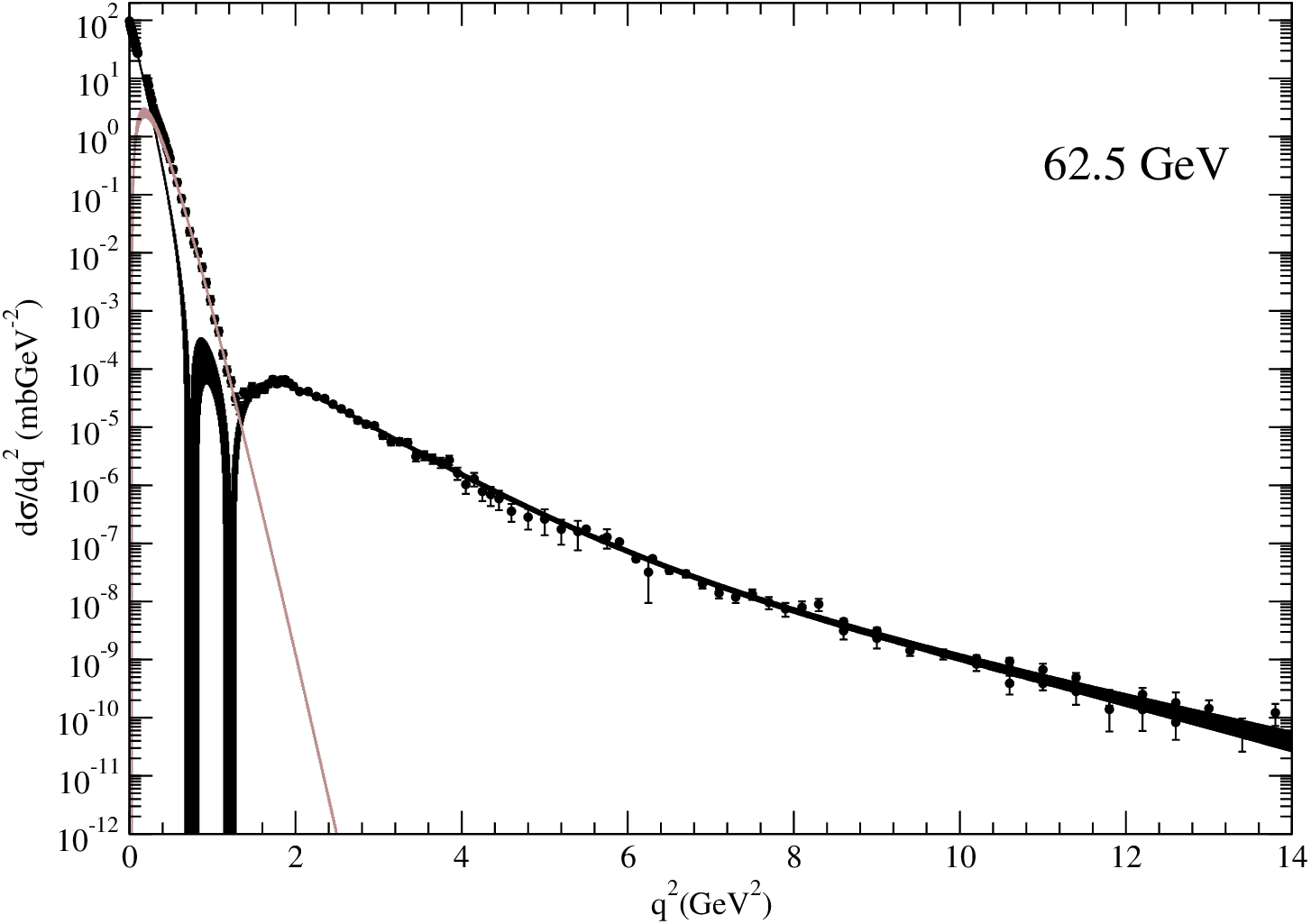}}
\caption{Method 2: uncertainty regions for the contributions to the differential cross sections 
from Re$\,A(s,q)$ (grey) and Im$\,A(s,q)$ (black) and the experimental data.}
\label{fig:4} 
\end{figure*}

\section{Discussion}
\label{sec:5}

Table 4 summarizes the statistical results obtained here and those from previous analyses, for comparison.
From that table, in terms of the $\chi^2$/DOF (closest to 1) and taking into account all the six 
sets analyzed, we see that Method 1 presents the best statistical result.
Except for the cases at 23.5 and 30.7 GeV the results with Method 2
are also better than those obtained in the previous analyses.
In this section we discuss the implication of the results 
with focus on the contributions  to the differential cross sections from the real and imaginary 
parts of the amplitude (Sect. 5.1) and an outlining of the impact parameter picture associated with the
overlap functions (Sect. 5.2). Based on this discussion and some
other aspects we shall select Method 1 as our
best result (Sect. 5.3) and then present some critical remarks (Sect. 5.4).

\subsection{Scattering amplitude}

The differential cross section data analyzed here are characterized by the well
known diffractive pattern: the diffraction peak, the dip region and the
small monotonic decrease at large values of the momentum transfer. It is generally believed
that this pattern is associated with a dominant imaginary amplitude, which is
supposed to vanish only at the dip position, where the real part dominates
filling up the dip.
This standard picture seems to be inspired in the small value of the $\rho$ parameter 
($q^2$ = 0) and in the fact that,
at the energy region investigated, the $\rho$ value goes through zero at the energies where
the dip is sharpest \cite{castaldi}. This view seems also to be corroborated by
Martin's theorem which states that the real part of the amplitude changes sign
at small values of the momentum transfer \cite{martin}, suggesting an effective contribution
only in the dip region. 

The above mentioned standard picture is in complete agreement with the results of our previous 
constrained analyses
\cite{am,cmm} and also with the predictions from the  great majority of the phenomenological models \cite{fiore}.
However that is not exactly the case with the results presented here, as summarized in what follows.

First we note that, from Tables 2 and 3, the numbers of exponential terms in Re$\,A$ and Im$\,A$ are rather different
with Methods 1 and 2. In the former case Re $A$ has 2 terms and Im $A$ has 6, and in the latter case, Re $A$ has 3 terms (except
at 30.7 and 44.7 with 2) and Im $A$ has 4 terms.
From Figures 3 and 4, the uncertainty regions in the 
contributions from Re $A$ and Im $A$ are 
much larger with Method 1 than with Method 2. Taking into account these uncertainties
the following features can be inferred in each case.

- Method 1.

Re$\,A$ is maximum at $q^2$ = 0, decreases monotonically and presents a zero
(change of sign) in the region 1--2 GeV$^2$. Im$\,A$ has, in general,  two zeros, one near the dip
and the other around 4 GeV$^2$, except at 62.5 GeV, where the second zero is not
present. Im $A$ dominates the peak and the region of large momentum transfer
and Re $A$ dominates the intermediate region, between the
dip ($\sim$ 1.2 GeV$^2$) and $q^2$ $\sim$ 6 - 7 GeV$^2$.

- Method 2.

Re $A$ increases from  $q^2$ = 0 up to a maximum around 0.3 GeV$^2$
and then decreases to zero just above the dip position (without change of sign).
Im $A$ presents two zeros, the first around 0.5 GeV$^2$ and the second at the dip position.
Im $A$ dominates the very forward peak and the region above the dip and
Re $A$ the peak.

Indeed, our results with Methods 1 and 2 are rather different from the above
mentioned standard picture. Before discuss our conclusions (Sect. 5.3), let us outline some
extracted results in the impact parameter space from Methods 1 and 2.

\begin{table}
\begin{center}
\caption{Values of the $\chi^2$ per degree of freedom (DOF) from previous
analysis by Carvalho--Martini--Menon (CMM), \'Avila--Menon (AM), and those obtained in this work
with Methods 1 and 2.}
\label{tab:4}
\begin{tabular}{ccccccc}
\hline
$\sqrt{s}$ (GeV): & \ 19.4 \ & \ 23.5\ &  \  30.7 \      & \ 44.7 \ & \ 52.8\  & \ 62.5\ \\
\hline
CMM \cite{cmm}: & 2.80   & 1.20  & 1.28   & 2.13   & 2.07   & 1.51 \\
AM \cite{am}:         & 2.76   & 1.20  & 1.24   & 2.05   & 1.71   & 1.22 \\
Method 1:  & 2.62   &1.09   & 1.09   & 1.99   & 1.55   & 1.18 \\
Method 2: & 2.53 & 1.77& 3.79&1.87 & 1.55& 1.18\\
\hline
\end{tabular}
\end{center}
\end{table}

\subsection{Impact picture}

By inverting Eq. (5), we extract the profile function and then the total,
elastic and inelastic overlap functions, Eqs. (7) and (8), together with the 
corresponding uncertainty regions. Typical results obtained with
Methods 1 and 2 are shown in Figure 5, in the
case of data at 52.8 GeV (largest set of experimental data). 
We see that despite all the differences in the contributions from Re $A$ and Im $A$
the impact pictures are similar: the two results agree to within 20 \%.

\begin{figure}[h]
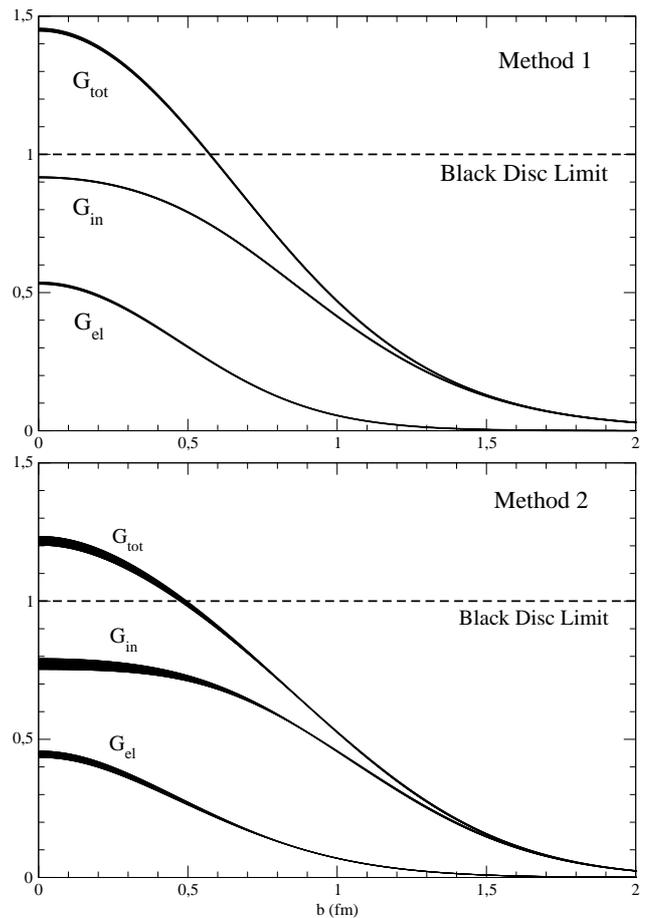

\resizebox{0.46\textwidth}{!}{\includegraphics*{f5a.eps}}
\resizebox{0.46\textwidth}{!}{\includegraphics*{f5b.eps}}
\caption{Extracted overlap functions, Eqs. (7) and (8), at 52.8 GeV, with the uncertainty regions
from Method 1 and Method 2.}
\label{fig:5} 
\end{figure}

Although not interested in the energy dependence of the free parameters,
it will be useful to investigate the dependence of extracted quantities on
the energy. However, it should be noted that the interval is relatively
small, 19.4--62.5 GeV and given the smooth variation of the
physical quantities in this interval
(differential/total cross section and $\rho$),
the same smooth variation is expected on the extracted quantities. As illustration, we consider the ratio
of the inelastic overlap function between two different energies,

\begin{eqnarray}
R(b) \equiv  \frac{G_{in}(\sqrt{s_2}, b)}{G_{in}(\sqrt{s_1}, b)},
\end{eqnarray}
with $\sqrt{s_2}$ = 62.5 GeV and $\sqrt{s_1}$ = 19.4 GeV (investigated interval).
The results with Methods 1 and 2 are shown in Fig. 6, together with the uncertainty regions,
by error propagation.
Once more, despite the distinct contributions from Re A and Im A, we see that both methods predict
a (standard) peripheral increase of $G_{in}$, although faster with Method 2 than with Method 1.

\begin{figure}[h]
\resizebox{0.46\textwidth}{!}{\includegraphics*{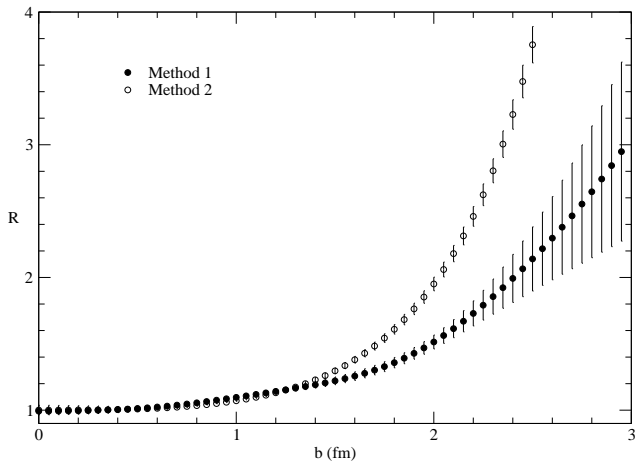}}
\caption{Ratio between the inelastic overlap functions,
Eq. (14), at 62.5 and 19.4
GeV, with Methods 1 and 2.}
\label{fig:6} 
\end{figure}

\subsection{Conclusion on the fit results}

As commented in the introduction, one of  our primary concerns is
to specify
what kind of solution we can arrive with some specific
methodology. We have considered two independent methods, based either
in previous results and excluding constraints (Method 1) or by testing all kind of possible contributions
in the real and imaginary parts of the amplitude and performing fits by regions of momentum transfer
and energy (Method 2). From Table 4, except for the results at 23.5 and 30.7 GeV with Method 2,
the reduced $\chi^2$ values are smaller and closer to 1 than those obtained in the previous analysis.
Besides that, the contributions from Re A and Im A are distinct with Methods 1 and 2 and both are rather different
from the standard picture. Concerning this last point we make the following observations.

- As explained in detail, we did not impose any theoretical idea or constraint in the
analytical parametrization or fit procedures either.

- Up to our knowledge, the standard picture (dominant Im A with Re A filling up the dip)
has never been proved or disproved. The arguments referred to in the beginning
of Sect. 5.1 seem to us plausible, but not conclusive.

-  We have found that some extracted quantities of interest (exemplified here with
$G_{in}(s,b)$) do not strongly differ with Method 1 or 2, and also from standard predictions of phenomenological models
(we shall return to this important point in Sect. 6).

Based on these observations, we understand that, although not standard, our empirical results
suggest another picture as possible, in both statistical and physical grounds.
However, we consider that Method 1 has led to the best results and that conclusion is based on the
following facts.

\begin{description}

\item[(1)] The reduced $\chi^2$ closest to 1 and smaller than that obtained in all our
previous analyses and in all energies investigated (Table 4). To our knowledge
that represents the best global statistical result in the literature, if only
statistical errors are considered.

\item[(2)] Smallest uncertainty regions in the directly extracted quantities, which is connected with the corresponding
variances, covariances and error propagation procedures (Figure 5 is a typical example).

\item[(3)] Re A presents a maximum at $q^2$ = 0 and its slope in this region is greater than that
presented by Im A, which is consistent with a recent model-independent
analysis by Kohara--Ferreira--Kodama on data at small momentum transfer \cite{kfk}.

\item[(4)] The very poor statistical result obtained with Method 2 at 23.5 and mainly 30.7 GeV,
as compared with all the other results (Table 4).

\item[(5)] The imaginary part of the profile function extracted with Method 2, at 
30.7 GeV and at all other energies, has opposite sign: a reflection through the 
impact parameter axis.

\end{description}

Method 2, described in Sect. 3.3.2, seems to us a fully unbiased procedure
and the above mentioned shortcomings in the final results were not expected. This
intrinsic unbiased character of the method and our goal to display what kind of 
solution can be obtained with a specific methodology were the reasons why we
have included details on the method and results in this work. Once selected Method 1 
as our best result let us discuss some physical implications and critical aspects 
involved.

\subsection{Critical remarks}

What is novel in our selected result (Method 1) is the dominance of Re A at the 
intermediated values of the momentum transfer (2 - 7 GeV$^2$) and also two changes 
of sign in Im A, except at 62.5 GeV (Figure 3). We note that the change of sign in 
Re A at small values of the momentum transfer is consistent with the prediction of 
the theorem by Martin \cite{martin}, but that behavior was not imposed in the 
parametrization as done in \cite{am,cmm}.

Despite the good fit quality with Method 1 (item (1) above), some 
comments are in order regarding the numerical values of the parameters at 
different energies, displayed in Table 2. First we note that the errors in 
the parameter $a_1$ are larger than the corresponding central values in all 
sets analyzed, which seems to violate the uncertainty criterion referred to 
in Sect. 3 (last paragraph). That, however, does not constitute an inconsistency 
since $a_1$ and also $c_1$ are not fit parameters: they have been eliminated 
through Eqs. (2) and (3) (note that the sum in parametrization (12) starts at 
$i, j = 2$) and  have been included in Table 2 only for completeness. This 
elimination also explains, in part, the apparently anomalous value of $c_1$ at 
44.7 GeV, as compared with the values at all the other energies. However, at 
44.7 GeV, not only the value of $c_1$ but also of $c_2$ and $c_5$ seem anomalous 
and we understand this effect as associated with the lack of experimental data 
at the diffraction peak, in the region $0.3  < q^2 < 0.5$ GeV$^2$, which does 
not occur in the other sets (except at 52.8 GeV in the region 0.45--0.65 GeV$^2$). 
That can also explain the high value obtained for the reduced $\chi^2$ at 44.7 GeV 
(compared with the other ISR sets), in all our analyses (present and previous), 
as can be seen in Table 4. 

The above mentioned and some others apparently anomalous values in 
different sets (for example $c_3$ and $c_5$) are also associated with our fit 
procedure and strategies: (1) method 1 was based in two different choices for 
the initial values of the free parameters (variants 1 and 2 referred to in Sect. 
3.3.2) and the selection of the result was based only on the reduced $\chi^2$ 
value, independently of the variant. That was possible because we did not identify 
any anomalous variation in the extracted quantities at the nearby energies; 
(2) with the Minuit code we did not impose any strategy in the search method, 
except for the starting values and steps, and no limits have been imposed to any 
parameter, since we are looking for a completely unbiased procedure.

On the other hand, it may be interesting to note that the above discussion 
applies only to the parameters associated with the imaginary part of the amplitude 
and not to those in the real part. In fact, from Table 2, all the values of the parameter
$b_1$ are of the same order of magnitude, lying approximately in the interval
0.76--0.93 GeV$^{-2}$ and the parameters $a_2$ and $b_2$ 
present an increase with the energy in all the interval investigated. 
Despite these regularities, we stress that we do not intend here to investigate any 
energy dependence in the free parameters; our central point is to get the best 
result on statistical grounds.

A distinct level of energy dependence concerns the extracted quantities 
and in that case, our results have, presently, limited applicability. The main reason 
is the relative small energy interval investigated ($\approx$ 19--60 GeV), which 
also corresponds to the region just above the point where $\rho$ changes sign and 
$\sigma_{tot}$ begins to increase, leading therefore, to several different possible 
uncontrolled extrapolations. As a consequence of this reduced interval, 
the uncertainty regions, evaluated through error propagation from the fit parameters, 
generally  overlap and do not allow detailed inferences on energy dependence at 
nearby sets. In this respect, the new data on the differential cross section from 
the LHC by the TOTEM collaboration, reaching the large momentum transfer region, 
will certainly shed light on this problem.

On the other hand, even with these limited conditions, if we consider only 
the extreme energies of the interval investigated, some trends can be inferred.
As illustration we plot in Figure 7 the uncertainty regions associated with 
the real and imaginary parts of the profile function, at 19.4 and 62.5 GeV, obtained 
with our selected Method 1. Taking into account these uncertainties, we note a 
peripheral increase in Re $\Gamma$ and an increasing contribution in Im $\Gamma$ 
through negative values, as the energy increases. In this interval both quantities 
present a change of curvature around 0.5 GeV, suggesting a change in the 
geometrical/optical structure, as predicted in the standard phenomenological 
context \cite{fiore}.

\begin{figure}[h]
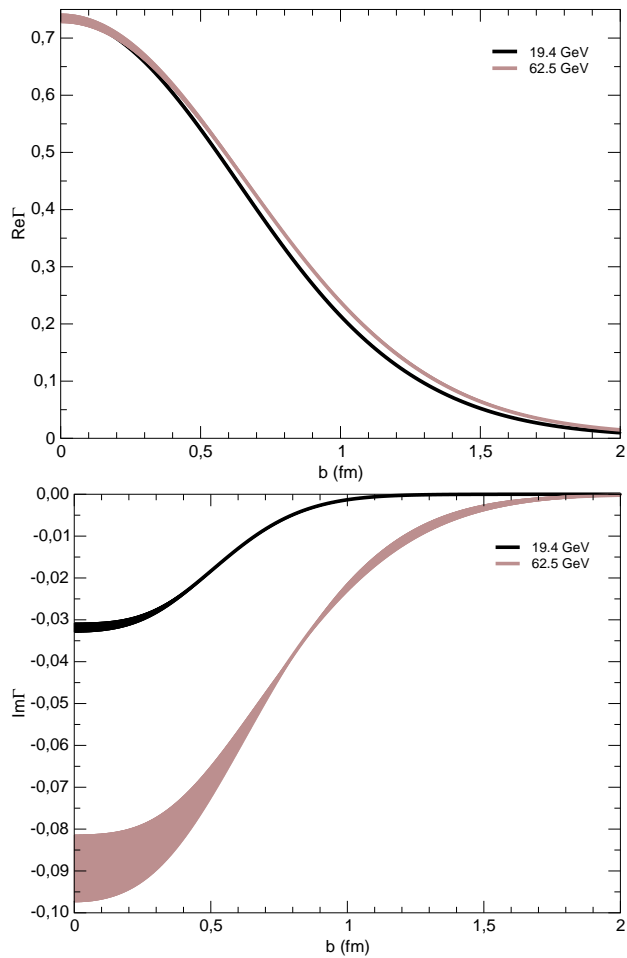

\resizebox{0.45\textwidth}{!}{\includegraphics*{f7a.eps}}
\resizebox{0.45\textwidth}{!}{\includegraphics*{f7b.eps}}
\caption{Real and imaginary parts of the profile function at
19.4 and 62.5 GeV, with Method 1.}
\label{fig:7} 
\end{figure}

We also note that if the new LHC data allow one to infer an energy dependence 
in our parameters, a predictive model-independent approach could be obtained, but 
that would demand the energy dependencies on $\sigma_{tot}(s)$ and $\rho(s)$, 
since they are input quantities in our analysis. However, that can be obtained, 
for example, from the detailed parametrization by the COMPETE Collaboration 
\cite{compete}.

As commented in our introduction, some aspects of the inverse problem can 
only be treated through detailed analyses of different solutions, connected with 
different parametrization and fit procedures. However, some characteristics can 
already be extracted from our analyses, since they are all based on the inclusion 
of the data at 27.4 GeV (large momentum transfer region) in the ISR sets, as 
explained in Sect. 3.1.1. As a consequence, all the empirical results in these 
analyses indicate a smooth monotonic decrease of the differential cross section 
at high values of the momentum transfer, that is without any oscillatory or
shoulder effect. Only to quote two typical examples, this result for $pp$ 
scattering at 52.8 GeV favors the three pomeron model by Petrov and Prokudin 
\cite{pp}, but not the hybrid model by Martynov and Nicolescu \cite{mn}.

At last, some comments on the data sets presently available and those used 
here are in order. All our analyses have been based on the data sets published by
experimental collaborations; the complete list of references are given in Sect. 
3.1.2 for the data at 19.4 GeV and in \cite{am} for the ISR sets (Landolt--B\"ornstein 
Series). Except for the normalization factor of the Faissler data at 19.4 GeV, 
we did not include the systematic errors in the analysis, only the statistical ones, 
as has been justified in Sect. 3.3.1. Moreover, we did not make any kind of data 
selection, but included all the published results. For us, that represents a 
criterion that allows comparisons among different parametrization, fit procedures 
and results, all based on the same data set. On the other hand, in the last years, 
some phenomenologists have argued that the raw published data need selections and 
also analyses on the systematic errors involved. Some compilations already exists, 
as the data set elaborated by Cudell, Lengyel and Martynov, discussed in \cite{clm} 
and available at \cite{cudell}. We think it would be interesting to develop new data 
reductions with this data set and all our empirical parametrization. A comparative 
analysis on all the results may be useful (see comments on this respect in \cite{mn}).

\section{Final conclusions and outlooks}
\label{sec:6}

The lack of a pure QCD description of the elastic scattering data and the distinct physical pictures associated
with the broad variety of phenomenological models, have been our primary motivation
for exploring the inverse problem, as a source of empirical information, suitable for model
developments and possible connections with QCD. In the context of an unitarized scheme (eikonal
representation), results from previous analyses have already demonstrated the feasibility of this
strategy, mainly related to empirical information on the eikonal in the
momentum transfer space \cite{am,cmm}.
Nevertheless, as discussed in detail, the main drawbacks concern the absence of a unique
solution, which is a consequence of the non-linear fit and the fact that the contributions from the real and
imaginary parts of the amplitude, beyond the forward direction, do not constitute physical observables. 

Here our focus was in the development of a full unconstrained parametrization, fits and procedures, stressing the kind of solution
that can be obtained with a specific methodology. In this context, despite the non standard picture 
related with the contributions from Re A and Im A (compared
with phenomenological models and previous analyses), we have selected Method 1 as our best result.
The novel feature is the dominance of the real part of the amplitude at intermediated values of the momentum
transfer.

At this point, the above mentioned drawbacks may suggest that we are faced with a messy situation, without 
conclusive directions/solutions and that the investigation of the inverse problem does not represent a way out.
However, the crucial point is to take these shortcomings as hints for developments and
improvements:
as commented in our introduction, we understand that an efficient
strategy demands detailed investigation of different analyses and results. Since Re A and Im A above $q^2$ = 0 do not constitute
physical observables, the main point is to look for empirical information that do not depend on, or is not
strongly connected with, these contributions. In that sense, the different results concerning our previous analyses (one zero in Re A and one zero in Im A) and those here selected (one zero in Re A and two in Im A), represent only first steps in that direction. 

Summing up our strategies and outlooks are three fold:

\begin{description}

\item[1.] Global comparative analysis of model-independent pa\-ram\-e\-tri\-za\-tion
and fit procedures, with focus on good statistical results and distinguished by
different contributions from Re A and Im A.

\item[2.] Taking into account the uncertainty regions from error propagation, selection and extraction of
empirical properties that are common to all the analyses, namely that are not (or not so strongly) connected
with specific contributions from Re A and Im A.

\item[3.] Use of these empirical information in the development of consistent phenomenological
models, looking for possible connections with QCD.

\end{description}

At last we observe that, in addition to the previous analysis \cite{am} and that presented here,
a third possibility has already been investigated through an almost model-independent
representation for Martin's real part formula \cite{fmartin} (without the scaling property).
This analysis has indicated two zeros in Re A and one in Im A \cite{fm10},
what is distinct from both our previous results and
those presented here. Moreover, empirical information
have been extracted which are consistent with the three analyses \cite{tesedani}.
The investigation along these lines are in progress and a global review 
on all our results are being prepared as a forthcoming
work.

\begin{acknowledgement}
We are thankful to J. Montanha for a critical reading of the manuscript.
This research was supported by FAPESP 
(Contracts No. 09/50180-0, No. 07/01938-1, No. 07/05953-5)
and CAPES, IFGW-UNICAMP.
\end{acknowledgement}

\end{document}